\documentclass[10pt]{iopart}
\usepackage{ae,aecompl}
\usepackage{color}
\usepackage{amssymb}
\usepackage{graphicx}
\usepackage{esint}
\usepackage{bm}
\usepackage{srcltx}

\begin{document}

\title{The role of short-range magnetic correlations in the gap opening
of topological Kondo insulators}
\author{E. Ramos$^1$, R. Franco$^1$, J. Silva-Valencia$^1$, M. E. Foglio$^2$
and M. S. Figueira$^3$}

\address{$^{1}$Departamento de F\'{\i}sica, Universidad Nacional de Colombia,
A. A. 5997, Bogot\'a, Colombia\\
$^2$Instituto de F\'{\i}sica ``Gleb Wataghin''
Universidade Estadual de Campinas, 13083-970 Campinas, S\~{a}o Paulo, Brazil\\
$^{3}$Instituto de F\'{i}sica, Universidade Federal Fluminense,
24210-340, Niter\'oi, RJ, Brazil}

\vspace{10pt}
\begin{indented}
\item[]June 2016
\end{indented}

\begin{abstract}
In this article we investigate the effects of short-range anti-ferromagnetic correlations on the gap opening of topological Kondo insulators. We add a Heisenberg term to the periodic Anderson model at the limit of strong correlations in order to allow a small
degree of hopping of the localized electrons between neighboring sites of the lattice.  This new model is adequate for studying   topological Kondo insulators, whose paradigmatic material is the compound $SmB_{6}$. The main finding of the article is that the short-range antiferromagnetic correlations present in some Kondo insulators contribute decisively to the opening of the Kondo gap in their density of states. These correlations are produced by the interaction between moments on the neighboring  sites of the lattice.

For simplicity, we solve the problem on a two dimensional square lattice. The starting point of the model is the $4f-Ce $ ions orbitals, with $J=5/2$ multiplet in the presence of spin-orbit coupling. We present results for the Kondo and for the antiferromagnetic correlation functions. We  calculate the phase diagram of the model, and as we vary the $E_{f}$ level position from the empty regime to the Kondo regime, the system develops  metallic and topological Kondo insulator phases. The band structure calculated shows that the model describes a strong topological insulator.
\end{abstract}

\vspace{2pc}
\submitto{\JPCM}
\maketitle
\ioptwocol

\section{Introduction}

\label{sec1}

The behavior of heavy fermion materials is characterized by the competition
between the Kondo effect, which tends to prevent a magnetic order in the system,
and Ruderman-Kittel-Kasuya-Yosida (RKKY) interaction, which tends to
magnetize the system. To describe the properties of these
systems, it is necessary to take into account these two tendencies on an equal
footing. It turns out that some cerium compounds such as $CeAl_{3}$, $%
CeCu_{6}$, and $CeRu_{2}Si_{2}$ are non-magnetic at very low temperatures and
exhibit a Fermi liquid behavior, while other compounds, such as $CeAl_{2}$, $%
CeB_{6}$, and $CeIn_{3}$ exhibit an antiferromagnetic order at low temperatures. This competition 
in heavy fermion systems has been previously studied by
Doniach \cite{Doniach1977} and after that by Coqblin {\it et al.} \cite%
{Coqblin96,Coqblin97}, who introduced a generalization of the Kondo
Hamiltonian that takes into account the Kondo effect and the short-range
magnetic correlations (SRMC) on an equal footing. They solved the Hamiltonian
using a technique similar to the slave-boson mean field theory (SBMFT) \cite%
{Coleman84}. 

Kondo insulators (KI) have been studied intensively
since their classification as highly correlated insulator
systems by Aeppli and Fisk \cite{Aeppli92}. Among the large
number of metallic rare earth compounds,
there are some that exhibit insulator behavior, as for example:
$Ce_{3}Bi_{4}Pt_{3}$, $SmB_{6}$,  and $YbB_{12}$. These materials exhibit a very small gap originating from localized (f-electrons) and conduction (c-electrons) electron hybridization, and at high temperatures,  $Ce$, $Sm$ and $Yb$ ions  exhibit  local moments  and share  their high-temperature properties with Kondo metals. The magnetic susceptibility  obeys the classic Curie-Weiss law  $\chi \approx M^{2}n_{i}/3(T+\Theta)$ \cite{Coleman15}, with $M$ being the magnetic moment, $n_{i}$ the magnetic moment concentration, and 
$\Theta$ the Curie-Weiss temperature, a phenomenological scale that takes into account the interactions between magnetic moments.

Our main aim in this paper is to study a novel class of $4f$ and 
$5f$ orbital intermetallic materials: the topological Kondo insulators (TKI), of
which $SmB_{6}$ is the first example \cite{ZFisk12, ZFisk13}.
Experimental measurements on the resistivity and susceptibility of doped $SmB_{6}$,  
detected that the transport and spin gaps exhibit approximately the same energy interval,
$E_{g} \simeq (10-10.5) \hspace{1.0mm} meV$, and develop below $70 \hspace{1.0mm} K$, 
due to the strong correlation between localized $4f$ states and the conduction 
$5d$ band. This strong interaction gives rise to a bulk insulating state at low
temperatures, while the surface remains metallic. This effect arises due to the
inversion of even parity conduction bands and odd parity localized electron
bands. 

We are interested in investigating the role of the short-range antiferromagnetic
correlations (SRAFC) in the gap opening of the topological Kondo insulators. These
correlations are produced by the interaction between moments on neighboring  
sites of the lattice. There are some experimental results obtained by
inelastic neutron scattering (INS) of some Kondo insulators: $YbB_{12}$ \cite%
{Mignot05}, $SmB_6$ \cite{Alekseev09}, and $CeNiSn$ \cite{Park98,Sato05}
that support the existence of such correlations in these systems. At low temperatures, the INS
spectra typically exhibit, response peaks within the
interval $[1-20] \hspace{1.0mm} meV$, which seem to be directly related to the low energy spin-gap
structure of the compound, and which disappear as the temperature is
increased. The SRAFC lead to the formation of
low energy peak structures at around $\hbar \omega=10 \hspace{1.0mm} meV$ for $YbB_{12}$
\cite{Mignot05}, at $\hbar \omega=14.5 \hspace{1.0mm} meV$ for $SmB_{6}$\cite%
{Alekseev09}, and at $\hbar \omega=2 \hspace{1.0mm} meV$ and $\hbar \omega=4 \hspace{1.0mm} meV$ for $%
CeNiSn $\cite{Sato05}, corresponding to different directions.

Another strong piece of evidence of the existence of the SRMC in some Kondo
insulators is provided by high pressure experiments \cite{Barla05a,Barla05b}. Resistivity measurements on $SmB_{6}$ performed by J. Derr {\it et al.} \cite{Derr08}  under optimum hydrostatic 
conditions, employing a diamond anvil cell with
argon as a pressure medium, showed that the insulating state vanishes, due to the closing of the topological Kondo gap (TKG), at a pressure of around $P=10 \hspace{1.0mm} GPa$, where a homogeneous long range magnetic order appears. The magnetic ordering temperature $T_{M}$ corresponds to a minimum of the resistivity $\rho$ and may mark an antiferromagnetic ground state with zone boundary reconstruction or nesting
effects. They also obtained the phase diagram of the system. In a recent article \cite{Paglione16}, measurements of the pressure dependence of the $SmB_{6}$ indicate that the material maintains a stable intermediate valence character up to a pressure of at least $35$ GPa, and the closure of the resistive activation energy gap and onset of magnetic order at $P=10 \hspace{1.0mm} GPa$ are not driven by stabilization of an integer valence state. This unexpected results seems to indicate that the compound supports a non trivial band structure. 

In the case of the $SmB_{6}$, the spin-orbit interaction lifts the
f-level multiplet degeneracy, generating two levels: the ground state with $%
J=5/2$ and an excited state with $J=7/2$. Generally, the state $J=7/2$ is not
considered, nor any boron state ($B_{6}$), because \textit{ab initio}
calculations \cite{FengLu13} indicate that they are far away from
the Fermi level. Now considering the ground state J=5/2, the crystal field
corrections due to  Samarium ions, split this state according to the irreducible representations of the cubic group $O_{h}$ into two degenerate states: $\Gamma_{8}(4)$ (ground state) and 
$\Gamma_{7}(2)$ (excited state), where
the number in parenthesis represents the degeneracy of the level. The 
$\Gamma_{8}$ states of the quartet have lobes along the axial directions: $%
\Gamma^{(1)}_{8}(2)$ along the $x$ and $y$-axes, and $\Gamma^{(2)}_{8}(2)$
along the $z$-axis \cite{Gong15}.

In this paper, we only consider the two-fold bi-dimensional irreducible representation 
$\Gamma^{(1)}_{8}$, which produces a V-shaped density of states in the TKI regime, and is a Kramers doublet whose form factor $[\mathbf{\Phi}]_{\sigma \alpha}(\mathbf{k})$ can be represented by a $2$x$2$
matrix. This represents the hybridization between the conduction electrons
characterized by the label $\sigma$ and the localized electrons
characterized by the label $\alpha=\pm$.  It is important to stress that the formalism developed here is sufficiently general to be applied to the ``minimum model" defined in the paper by Dzero {\it et al.}\cite{Dzero13,Vojta15} to describe the low temperature physics of the $SmB_{6}$,
in which the $\Gamma_{8}$ quartet hybridizes with the $e_{g}$ quartet
(Kramers doublet plus spin degeneracy).

To solve the full problem, we must include the $\Gamma^{(2)}_{8}$ irreducible
representation in the calculations for a tight binding three-dimensional lattice. The generalization to this case is straightforward,  but the numerical computation cost increases greatly, mainly in the calculation of the density of states. We do not present such calculations here because the focus of this
paper is the study of the role of the SRAFC in the gap formation of some Kondo insulators. However, the results obtained in the present paper should be relevant to the study the Kondo insulator $CeNiSn$, which
exhibits a spin gap originating from SRAFC \cite{Park98,Sato05} and a
V-shaped density of states \cite{Ikeda96,Nakamura96,JuanaMoreno2000}.

This paper is organized as follows: in Sec. \ref{sec2}, we define the periodic Anderson model in terms of the $X$ Hubbard operators and present the calculation of the Green's functions. In Sec. \ref{sec3}, we discuss the basic theory of the X-boson approach and the calculation of related parameters. In Sec. \ref{sec4},  we discuss the results and their physical consequences. In Sec. \ref{sec5} we summarize the results and present the concluding remarks. Finally, in \ref{sec6} we develop the mean field calculation  of the Heisenberg Hamiltonian employed in the paper and in \ref{sec7} we discuss some details of the calculation of the cumulant Green's functions with the inclusion of the direct hopping between nearest-neighbors sites of the lattice.

\section{The periodic Anderson model}
\label{sec2}

TKI have been studied employing slave boson mean field theory
(SBMFT) \cite{Coleman84}, at the limit of infinite Coulomb repulsion $%
U\rightarrow \infty $ \cite{Dzero12,Tran12} and for finite correlation $U$
\cite{Sigrist14}. SBMFT  is less adequate for describing
intermediate valence (IV) systems  like the new TKI $SmB_{6}$, due to the presence  of an unphysical temperature second order phase transition where  conduction and  localized electrons decouple from each other. To circumvent these problems, maintaining the simplicity of the calculation
and the ideas involved in SBMFT, we generalize our previous work on the $X-$boson 
approach \cite{Xboson} to the periodic Anderson model (PAM), considering $f$ electrons states with a
total angular momentum $J$ and $z$-axis component $M$, while the conduction
electron states are described by momentum $\mathbf{k}$ and spin $\sigma$. The spin-orbit coupled
Wannier states of the conduction electrons are then decomposed in terms of
plane-wave states, and give rise to a momentum-dependent hybridization
characterized by form factors with symmetries that are uniquely determined
by the local symmetry of the $f$ states  \cite{Tran12}.

The X-boson approach to the periodic Anderson model \cite{Xboson} at the
limit of infinite Coulomb correlations ($U=\infty $) has been already
studied, employing the cumulant expansion \cite{Hubbard5,FFM}. In these papers,
the Hubbard operators $X_{j,ab}$=$\left\vert j,a\right\rangle \left\langle
j,b\right\vert $ were employed, where the set $\left\{ \left\vert
j,a\right\rangle \right\} $ is an orthonormal basis in the space of
interest. Projecting out the components with more than one electron from any
local state at site $j$, one obtains
\begin{equation}
H=H_{o}+H_{h}  \label{Hfull}
\end{equation}
with
\begin{equation}
H_{o}=H_{c}+H_{f}^{o} ,  \label{HO}
\end{equation} 
where
\begin{equation}
H_{c}=\sum_{\mathbf{k}\sigma }E_{\mathbf{k},\sigma }c_{\mathbf{k},\sigma
}^{\dagger }c_{\mathbf{k},\sigma }  \label{Hconductio}
\end{equation}%
is the Hamiltonian of the conduction electrons ($c$-electrons), with
momentum $\vec{k}$ and spin $\sigma $, and
\begin{equation}
H_{f}^{o}=\sum_{j\alpha }\ E_{f,\alpha} X_{j,\alpha \alpha },
\label{Hlocal}
\end{equation}%
corresponds to independent localized electrons ($f$-electrons), with
pseudospin $\alpha $ belonging to the representation $\Gamma^{(1)}_{8} $ of the
multiplet state at the site $j$. The last term in Eq.(\ref{Hfull})
\begin{equation}
H_{h}=\sum_{j\alpha ,\mathbf{k}\sigma }\left( V_{j,\sigma \alpha }(\mathbf{k}%
)X_{j,\alpha }^{\dagger }c_{\mathbf{k},\sigma }+V_{j,\sigma \alpha }^{\ast }(%
\mathbf{k})c_{\mathbf{k},\sigma }^{\dagger }X_{j,\alpha }\right) ,
\label{Hhibrid}
\end{equation}%
is the hybridization Hamiltonian giving the interaction between the $c$-electrons and the $f$-electrons, 
with $V_{j,\sigma \alpha }(\mathbf{k})=(1/\sqrt{N_{s}})V_{\sigma \alpha }(\mathbf{k})\exp {(i\mathbf{k}.\mathbf{R}_{j})}$, 
where ${\mathbf{R}_{j}}$ is the position of site $j$ and $N_{s}$ is the
number of lattice sites. We should note that this interaction conserves the spin component 
$\sigma$. Since there is no local hybridization process between conduction ($s-$electrons) and 
localized ($f-$electrons) eletrons in rare earth ions, the
hybridization results from the nearest-neighbor hopping from the $f$
electrons at a site $j$ to the $s$ electrons in the vicinity of this site
\cite{Tran12}. We will consider the site $j$ independent hybridization.

Since the treatment employs the grand canonical ensemble, instead of $H$ {we
will use
\begin{equation}
\mathcal{H}=H-\mu \left\{ \sum_{\mathbf{k,\sigma }}c_{\mathbf{k,\sigma }%
}^{\dagger }c_{\mathbf{k,\sigma }}+\sum_{j\alpha}\nu _{\alpha}X_{j,\alpha\alpha}\right\} ,
\label{HGCE}
\end{equation}%
\noindent where $\nu _{a}=0,1$ is the number of electrons in state $\mid
a> $. It is then convenient to define
\begin{equation}
\varepsilon _{\mathbf{k,\sigma }}=E_{\mathbf{k,\sigma }}-\mu \ ,
\label{Eq2.4}
\end{equation}%
and}%
\begin{equation}
\varepsilon _{f\mathbf{,\alpha }}=E_{f,\alpha}-\mu 
\end{equation}%
{because $E_{\mathbf{k,\sigma }}$ and }$E_{f,\alpha}${\ appear
only in that form in all the calculations. }

\bigskip In order to take into account the short-range magnetic correlations between
neighboring f-electrons of the lattice, we include in Eq. \ref{HGCE} the Heisenberg Hamiltonian 
\begin{equation}
H_{f}^{\prime} =-J_{H}\sum\limits_{\left\langle i,j\right\rangle
}\mathbf{S}_{i} \cdot \mathbf{S}_{j} ,
\label{Hopp1}
\end{equation}
where $J_{H}$ represents the exchange integral and the operators $\mathbf{S}_{(i,j)}$ represent the nearest-neighbors  magnetic moments of the $i,j$ sites of the lattice.  When $J_{H}<0$ ($J_{H}>0$) the magnetic correlations favor  antiferromagnetism (ferromagnetism). Expressing the Heisenberg Hamiltonian in terms of Hubbard operators,  we show in \ref{sec6} that at a mean field level, this Hamiltonian can be put into the form
\begin{equation}
H_{f}^{\prime}=\sum_{i,j,\alpha}t_{i,j,\alpha}\ X_{i,0\alpha}^{\dagger}X_{j,0\alpha} ,
\label{E121}
\end{equation}
where 
\begin{equation}
t_{i,j,\alpha}=-\frac{1}{2}J_{H} \langle X^{\dagger}_{i\protect\alpha}X_{j\protect\alpha} \rangle, 
\label{E131}
\end{equation}
represents the correlated hopping of $f$ electrons between neighboring sites of the lattice and  
$t_{i,j,\alpha}=t_{j,i,\alpha}^{\ast }$ to satisfy the Hermitean character of the Hamiltonian. The function  
$\langle X^{\dagger}_{i\protect\alpha}X_{j\protect\alpha} \rangle$ is a quantity that measures the correlation between neighboring $f$-electrons of the lattice and  must be calculated self-consistently. 

The $X$-Hubbard {\ operators are adequate to work with local states associated
with the sites }$j$ of a lattice, and are defined in general by $X_{j,ab}$=$%
\left\vert j,a\right\rangle \left\langle j,b\right\vert $, where the set $%
\left\{ \left\vert j,a\right\rangle \right\} $ is an orthonormal basis in
the space of local states of interest. They do not satisfy the usual
commutation relations, and therefore the diagrammatic methods based on
Wick's theorem are not applicable. Instead of commutation relations, one has
the product rules $X_{j,ab}.X_{j,cd}=\delta _{b,c}X_{j,ad}$,  
and we will use a cumulant expansion that was originally employed by
Hubbard \cite{Hubbard5} to study his model. This expansion was later
extended to the PAM \cite{FFM}, but here we will have to make a further
extension in order to include the Heisenberg Hamiltonian $H_{f}^{\prime }$ in the perturbation $H_{1}$, 
\begin{equation}
H_{1}=H_{h}+H_{f}^{\prime }.  \label{Eq1.7}
\end{equation}%
When $U\rightarrow \infty$, the identity $I_{j}$ at site $j$ should satisfy
the completeness relation:
\begin{equation}
X_{j,00}+X_{j,(++)}+X_{j,(--)}=I_{j},  \label{Complete}
\end{equation}
where the first Hubbard operator represents the vacuum state and the last two, with the labels 
$(++)$ and $(--)$, represent the pseudospin components associated with the Kramer
doublet of the irreducible representation $\Gamma_{8}^{(1)}$. The
occupation numbers $n_{j,\alpha }=\langle X_{j,\alpha \alpha }\rangle$ can be calculated
from appropriate Green's functions (GF), and assuming translational
invariance we can write $n_{j,\alpha }=n_{\alpha }$ (independent of the site $j$), so
that we can write
\begin{equation}
n_{o}+n_{+}+n_{-}=1.  \label{Eq1.9}
\end{equation}

In a way similar to SBMFT \cite{Coleman84},
the X-boson approach consists of adding the product of each Eq. (\ref%
{Complete}) times a Lagrange multiplier $\Lambda _{j}$ to Eq.~(\ref{HGCE}),
and the new Hamiltonian generates the functional that we will minimize
employing Lagrange's method. We introduce the parameter
\begin{equation}
R\equiv \langle X_{oo} \rangle,  \label{Eq.5}
\end{equation}
and we call the method ``X-boson'' because the Hubbard operator $X_{j,oo}$
has a ``Bose-like'' character \cite{FFM}, but note that  we do
not write any $X$ operator as a product of ordinary Fermi or Bose operators
as in SBMFT, but retain them in their original form. Imposing the
completeness relation (Eq. \ref{Complete}) on the full Hamiltonian given by 
Eqs. \ref{HGCE} and \ref{E121}, and introducing the Lagrange multiplier $\Lambda _{j}=\Lambda $, we obtain a new Hamiltonian
\begin{eqnarray}
&\mathcal{H}=\sum_{\mathbf{k}\sigma }\epsilon _{\mathbf{k},\sigma }c_{%
\mathbf{k},\sigma }^{\dagger }c_{\mathbf{k},\sigma }+\sum_{j\alpha }\ \tilde{%
\varepsilon}_{f,\alpha }X_{j,\alpha \alpha }  \nonumber \\
&+\sum_{j\alpha ,\mathbf{k}\sigma }\left( V_{\sigma \alpha }(\mathbf{k}%
)X_{j,\alpha }^{\dagger }c_{\mathbf{k},\sigma }+V_{\sigma \alpha }^{\ast }(%
\mathbf{k})c_{\mathbf{k},\sigma }^{\dagger }X_{j,\alpha }\right)  \nonumber \\
&+\sum_{i,j,\alpha }t_{i,j,\alpha}\ X_{i,\alpha }^{\dagger }X_{j,\alpha
}+N_{s}\Lambda (R-1),  \label{HSB}
\end{eqnarray}%
\noindent but with a renormalized localized energy
\begin{equation}
\tilde{\varepsilon}_{f,\alpha }=\varepsilon _{f\mathbf{,\alpha }}+\Lambda .
\label{Energ}
\end{equation}

In  \ref{sec6}, we show that in the presence of  hybridization and 
hopping between nearest-neighbor $f$-electrons, the mean field Green's functions \cite{Xboson} are given by:

\begin{equation}
G_{\mathbf{k}{\sigma },\alpha }^{ff}(z_{n})=\frac{-D_{\alpha }\left(
z_{n}-\varepsilon _{\mathbf{k}\sigma }\right) }{\left( z_{n}-\varepsilon _{%
\mathbf{k}\alpha }^{f}\right) \left( z_{n}-\varepsilon _{\mathbf{k}\sigma
}\right) -|V_{\sigma \alpha }(\mathbf{k})|^{2}D_{\alpha }},  \label{Eq3.155}
\end{equation}%
\begin{equation}
G_{\mathbf{k}{\sigma },\alpha }^{cc}(z_{n})=\frac{-\left( z_{n}-\varepsilon
_{\mathbf{k}\alpha }^{f}\right) }{\left( z_{n}-\varepsilon _{\mathbf{k}%
\alpha }^{f}\right) \left( z_{n}-\varepsilon _{\mathbf{k}\sigma }\right)
-|V_{\sigma \alpha }(\mathbf{k})|^{2}D_{\alpha }},  \label{Eq3.166}
\end{equation}%
\begin{equation}
G_{\mathbf{k}{\sigma },\alpha }^{fc}(z_{n})=\frac{-\ D_{0\alpha }V_{\sigma
\alpha }(\mathbf{k})}{\left( z_{n}-\varepsilon _{\mathbf{k}\alpha
}^{f}\right) \left( z_{n}-\varepsilon _{\mathbf{k}\sigma }\right)
-|V_{\sigma \alpha }(\mathbf{k})|^{2}D_{\alpha }},  \label{Eq3.177}
\end{equation}%
\noindent where we consider the  analytic continuation of the Matsubara frequencies $z_{n} \rightarrow \omega+i\eta$ to the real axis and $\varepsilon _{\mathbf{k}\alpha }^{f}=\tilde{\varepsilon}%
_{f,\alpha }+D_{\alpha }\overline{E}_{k\alpha}$ with $D_{\alpha }=\langle X_{oo} \rangle +
\langle X_{\alpha\alpha }\rangle$. In order to obtain explicit results for
the X-boson parameters, we will consider a spin independent tight-binding
conduction band  with hopping $t$ between nearest-neighbors
on a two-dimensional ($2D$) square lattice \cite{Werner13}
\begin{equation}
\overline{E}_{\mathbf{k\alpha}}=-\sum_{l}t_{i,i+l,\alpha}\exp [\mathbf{k}_{i}.\mathbf{R}%
_{l}] , \label{Eq40}
\end{equation}
where $\mathbf{k}=(\mathbf{k_{x}},\mathbf{k_{y}}).$ Considering the site $i$ as the origin and substituting Eq. \ref{E131} into Eq. \ref{Eq40} we obtain
\begin{equation}
\overline{E}_{\mathbf{k\alpha}}=-\frac{1}{2}J_{H} \langle X^{\dagger}_{i\protect\alpha}X_{j\protect\alpha} \rangle  \varepsilon _{\mathbf{k}} ,
\label{Eq401}
\end{equation}
with
\begin{equation}
\varepsilon_{\mathbf{k}}=-2t\sum_{i=x,y}[cos(k_{i}a)] - \mu , 
\label{TB}
\end{equation}
where $a$ is the lattice parameter and we also consider $t=t_{x}=t_{y}$.

\section{The X-boson approach}
\label{sec3} 

In this section, we extend the X-boson treatment \cite{FFM} to the model
studied in this paper, and we will also discuss the thermodynamic potential {%
$\Omega =-k_{B}T\ln (\mathcal{Q})$, where }$\mathcal{Q}${\ is the grand
partition function and $k_{B}$ is the Boltzmann constant. A convenient way
of calculating $\Omega $ is to employ the method of{\ }${\xi }$ parameter
integration \cite{Doniach}. This method introduces a $\xi $ dependent
Hamiltonian $H({\xi })=H_{o}+{\xi }H_{1}${\ \ through a coupling constant }${%
\xi }${\ (with $0\leq {\xi }\leq 1$), where $H_{1}$ is given by Eq. (\ref%
{Eq1.7}). For each }${\xi }$, there is an associated{\
thermodynamic potential $\Omega ({\xi })$ that satisfies: \
\begin{equation}
\left( \frac{\partial \Omega }{\partial {\xi }}\right) _{V_{s},T,\mu
}=\langle H_{1}({\xi }) \rangle_{{\xi }},  \label{Eqn.14}
\end{equation}
\noindent where $<A>_{{\xi }}$ is the ensemble average of the operator }$A$
for a{\ system with Hamiltonian $H({\xi })$ and the given values of chemical
potential $\mu ,$ temperature $T$, and }volume $V_{s}${. Integrating this
equation gives} }

\begin{equation}
\Omega =\Omega _{o}+\int_{0}^{1}d{\xi }\langle H_{1}({\xi }) \rangle_{{\xi }},
\label{Eqn.15}
\end{equation}%
\noindent where $\Omega _{o}$ is the thermodynamic potential of the system
with ${\xi }=0$. This value of ${\xi }$ corresponds to a system without
hybridization and without hopping of $f$-electrons. One obtains in the
absence of magnetic field 
\begin{eqnarray}
&\Omega _{o}=\left(\frac{-1}{\beta }\right){\sum_{\mathbf{k}\sigma}}\ln \left[ 1+e^{-\beta
\varepsilon_{\mathbf{k}\sigma}}\right] + \nonumber \\
&\left(\frac{-N_{s}}{\beta }\right) \ln \left[1+2e^{-\beta \tilde{\varepsilon}_{f\alpha}}\right] + 
N_{s}\Lambda (R-1).  \label{Eqn.16}
\end{eqnarray}

In the present paper, the perturbation has two different types of
contributions: the hybridization $H_{h}$ and the hopping of the $f$%
-electrons $H_{f}^{\prime }$. The average $\langle H_{h} \rangle_{{\xi }}$ and its
corresponding contribution to $\Omega $ have already been calculated in
reference \cite{Xboson} for the system without $f$-electron hopping.
Including the $H_{f}^{\prime }$ in the perturbation $H_{1}$ and following
the same technique employed in this reference, one obtains {\
\begin{eqnarray}
& \left\langle H_{1}\right\rangle _{{\xi }}=\frac{1}{\pi }%
\int\limits_{-\infty }^{\infty }d\omega \ n_{F}(\omega ) \sum_{\mathbf{k,}\alpha \sigma }Im  \nonumber \\
& \left[ \frac{{\xi }\
\overline{|V_{\sigma \alpha }(\mathbf{k})|}^{2}-\overline{E}_{\mathbf{k\alpha}%
}^{s}\left( \omega ^{+}-\varepsilon _{\mathbf{k}}\right) }{\left( \omega
^{+}-\varepsilon _{f}-\xi \ \overline{E}_{\mathbf{k\alpha}}^{s}\right) \left(
\omega ^{+}-\varepsilon _{\mathbf{k}}\right) -{\xi }^{2}\ \overline{%
|V_{\sigma \alpha }(\mathbf{k})|}^{2}}\right] ,  \label{Eq.18}
\end{eqnarray}
where }$n_{F}(x)=1/\left[ 1+e^{\beta x}\right] ${\ is the Fermi-Dirac
distribution} and $\omega ^{+}=\omega +i\eta$.

As in the case of the system with $H_{f}^{\prime }=0$,\ Eq. (\ref{Eq.18}){\
has} an interesting scaling property: it is equal to the corresponding
expression of the uncorrelated system for the scaled parameters $\overline{V}%
_{\sigma \alpha }(\mathbf{k})$ and $\overline{E_{\mathbf{k\alpha}}^{s}}$ (it is
enough to remember that by replacing $D_{\alpha }=1$ in the GF given by 
Eqs. \ref{Eq3.155}-\ref{Eq3.177}, one obtains the corresponding GF 
of the uncorrelated system). Rather than
performing the ${\xi }$ and $\omega $\ integrations, we will use the value
of the $\Omega ^{u}$ for the uncorrelated system with $\overline{V}_{\sigma
\alpha }(\mathbf{k})=\sqrt{D_{\alpha }}V_{\sigma \alpha }(\mathbf{k})$ and $%
\overline{E_{\mathbf{k\alpha}}^{s}}=\overline{E}_{\mathbf{k\alpha}}\ {D_{\alpha }}$, and
then employ Eq. (\ref{Eqn.15}) to calculate 

\begin{equation}
\int_{0}^{1}d{\xi }\langle H_{1}^{u}({\xi }) \rangle_{{\xi }}=\Omega ^{u}-\Omega _{o}^{u} ,
\label{Omegadif}
\end{equation}  
where considering  $\overline{V}_{j,\mathbf{k},\sigma }=\overline{E_{\mathbf{k\alpha}}}=0$, we can write
\begin{eqnarray}
&\Omega_{o}^{u}=\left(\frac{-1}{\beta }\right){\sum_{\mathbf{k\sigma}}}\ln \left[ 1+e^{-\beta
\varepsilon _{\mathbf{k\sigma}}}\right] + \nonumber \\
&\left(\frac{-N_{s}}{\beta }\right)  \ln \left[ 1+e^{-\beta \tilde{\varepsilon}_{f\alpha}}\right]^{2} + 
N_{s}\Lambda (R-1)  \label{Eqn.19} .
\end{eqnarray}

Substituting Eq. \ref{Omegadif} into Eq. \ref{Eqn.15}, we can write

\begin{equation}
\Omega =\overline{\Omega }_{0} + \Omega ^{u} ,
\label{Omega1}
\end{equation}
where
\begin{equation}
\overline{\Omega }_{0}\equiv \Omega _{o}-\Omega _{o}^{u}=\left(-\frac{N_{s}}{%
\beta }\right)\ln \left\{ \frac{1+2\exp (-\beta \tilde{\varepsilon}_{f\alpha})}{\left[
1+\exp (-\beta \tilde{\varepsilon}_{f\alpha})\right] ^{2}}\right\} .  \label{Eqn.24}
\end{equation}

In our case, the unperturbed Hamiltonian for the lattice problem is
\begin{eqnarray}
&H^{u}=\sum_{\mathbf{k}\sigma }\ \varepsilon _{\mathbf{k},\sigma }\ c_{%
\mathbf{k},\sigma }^{\dagger }c_{\mathbf{k},\sigma }+\sum_{\mathbf{k}\alpha
}\varepsilon _{\mathbf{k}\alpha }^{f}\ f_{\mathbf{k},\alpha }^{\dagger }f_{%
\mathbf{k},\alpha } + \nonumber \\
& \sum_{\mathbf{k}\sigma,\alpha}\left( \overline{V}_{\sigma \alpha }(%
\mathbf{k})\ f_{\mathbf{k},\alpha }^{\dagger }c_{\mathbf{k},\sigma }+%
\overline{V}_{\sigma \alpha }^{\ast }(\mathbf{k})c_{\mathbf{k},\sigma
}^{\dagger }f_{\mathbf{k},\alpha }\right) + \nonumber \\
& N_{s}\Lambda (R-1) ,  \label{Eqn.20}
\end{eqnarray}
where $\varepsilon _{\mathbf{k}\alpha }^{f}=\tilde{\varepsilon}_{f,\alpha}+
D_{\alpha }\overline{E}_{k\alpha}$, $\tilde{\varepsilon}_{f,\alpha }=E_{f,\alpha}^{o}+
\Lambda -\mu$ and  $\overline{E}_{k\alpha}$ is given by Eqs. (\ref{Eq401}) and (\ref{TB}).

\bigskip {This Hamiltonian can be easily diagonalized, and  the corresponding
}$\mathcal{H}^{u}$ can be written as
\begin{equation}
\mathcal{H}^{u}=\sum_{\mathbf{k}\sigma,\alpha}\omega _{\mathbf{k}\sigma,\alpha}
v_{\mathbf{k}\sigma,\alpha }^{\dagger }v_{\mathbf{k}\sigma,\alpha} + N_{s}\Lambda (R-1)
\ ,  \label{Eqn.21}
\end{equation}%
where $v_{\mathbf{k}\sigma,\alpha }^{\dagger }$ ($v_{\mathbf{k}\sigma,\alpha }$) are the
creation (destruction) operators of the composite particles of energies $%
\omega _{\mathbf{k}\sigma,\alpha }$. The calculation of
\begin{eqnarray}
&\Omega ^{u}=\left( \frac{-1}{\beta }\right) \sum_{\mathbf{k}\sigma,\alpha }\ln \left[
1+\exp (-\beta \ \omega _{\mathbf{k}\sigma,\alpha})\right] + \nonumber \\
& N_{s}\Lambda (R-1) ,  \label{Eqn.22}
\end{eqnarray}
is straightforward, and substituting Eq. \ref{Eqn.22} into Eq. \ref{Omega1} we can write
$$
\Omega =\overline{\Omega }_{0}+\left(\frac{-1}{\beta}\right)\sum_{\mathbf{k}%
\sigma, \alpha, \ell =\pm }\ln \left[ 1+\exp (-\beta \ \omega _{\mathbf{k}%
\sigma,\alpha }(\ell )\right] +
$$
\begin{equation}
N_{s}\Lambda (R-1) . \label{Eqn.23}
\end{equation}

The Hamiltonian Eq. \ref{Eqn.20} can be written in a compact  form \cite{Tran12}
\begin{equation}
{\cal H}^{u}=\sum_{\bf{k}} \Psi^{\dagger }({\bf{k}}) H^{u} ({\bf {k}})\Psi({\bf{k}}) + N_{s} \Lambda (R-1),  \label{Eqn.21}
\end{equation}
with
\begin{equation}
H^{u}({\bf{k}}) = \left[ \begin{array}{cccc}
\varepsilon _{{\bf k}} & \overline{V}_{\sigma \alpha}  & 0 & 0 \\
\overline{V}^{\ast}_{\sigma \alpha} & \varepsilon _{\mathbf{k}\alpha }^{f} & 0 & 0 \\
0 & 0 & \varepsilon _{{\bf k}} & -\overline{V}_{\sigma \alpha}   \\
0 & 0 & -\overline{V}^{\ast}_{\sigma \alpha} &  \varepsilon _{\mathbf{k}\alpha }^{f} \end{array} \right] ,
\label{HD1}
\end{equation}
where $\Psi^{\dagger }({\bf{k}})=(c_{{\bf k},\uparrow}^{\dagger} f_{{\bf k},(-) }^{\dagger}c^{\dagger}_{{\bf k},\downarrow} f^{\dagger}_{{\bf k},(+)})$ is a four component Dirac spinor. The hybridization exhibits a more complex $\mathbf{k}$ dependence 
$\overline{V}_{\sigma \alpha} ({\bf k})=\sqrt{D_{\sigma}} V \mathbf{\Phi }(\mathbf{k})$ where
$|V_{\sigma \alpha} ({\bf k})|^{2}= D_{\sigma}|V|^{2} \mathbf{\Delta}^{2}(\mathbf{k})$ with
\begin{equation} 
\mathbf{\Delta }^{2}(\mathbf{k})=\frac{1}{2}Tr[\mathbf{\Phi}^{\dagger}(\mathbf{k}).\mathbf{\Phi }(\mathbf{k})] , 
\label{Form}
\end{equation}
where $[\mathbf{\Phi }]_{\sigma \alpha }(\mathbf{k})$, is the form factor, 
which is associated with the $\mathbf{k}$ dependence and the non trivial orbital structure of the
hybridization. Following the derivation presented in reference \cite{Tran12}, the form factor can be written as \cite{Werner13} $\mathbf{\Phi }(\mathbf{k})=\mathbf{d}(\mathbf{k})\circ {\bm{\sigma}}$, with ${\bm{\sigma}}$ being the Pauli spin matrices. For the 
bi-dimensional irreducible representation $\Gamma_{8}^{(1)}$, $\mathbf{d}(\mathbf{k})=2[sin(k_{x}),sin(k_{y})]$,
and the result for the $\mathbf{k}$ dependent hybridization function is
\begin{equation}
\mathbf{\Delta }^{2}(\mathbf{k})=4[sin^{2}(k_{x})+sin^{2}(k_{y})].
\label{Hyb}
\end{equation}

The eigenvalues $\omega _{\mathbf{k}\sigma,\alpha }$ of the{%
\ \ }$\mathcal{H}^{u}$ are just given by the poles of the GF in the mean field equations:  
Eqs.~(\ref{Eq3.155} - \ref{Eq3.177}). Due to the conservation of $\mathbf{k}$, the
Hamiltonian is reduced into $N_{s}$ matrices $2\times 2$ for each spin
component $\sigma $, and each pseudospin $\alpha $ belongs to the representation $\Gamma^{(1)}_{8}$, as indicated in the matricial Eq. \ref{HD1}. In this way, the 
$\omega _{\mathbf{k}\sigma, \alpha }$ can be calculated analytically
\begin{eqnarray}
&\omega_{\mathbf{k}\sigma, \alpha }^{\pm }= \frac{1}{2}\left(
\varepsilon_{\mathbf{k},\sigma }+\varepsilon _{k,\alpha }^{f}\right) \pm \nonumber \\
&\frac{1}{2}\sqrt{\left( \varepsilon _{\mathbf{k,}\sigma }-\varepsilon_{\mathbf{k,}%
\alpha }^{f}\right)^{2}+4 |V|^{2}D_{\alpha } 
\mathbf{\Delta}^{2}(\mathbf{k})} .\label{Eqn.32}
\end{eqnarray}

In an earlier paper \cite{Edwin2014}, we showed that expanding $\Phi _{%
\mathbf{k}}$ for small values of $\mathbf{k_{x}}$ and $\mathbf{k_{y}}$, we
obtain an effective Dirac theory given by the Hamiltonian
\begin{equation}
H^{u}(\mathbf{k})=\left[
\begin{array}{cccc}
\varepsilon _{\mathbf{k}} & \overline{V} & 0 & 0 \\
\overline{V}^{\ast} & \varepsilon _{\mathbf{k}\alpha }^{f} & 0 & 0 \\
0 & 0 & \varepsilon _{\mathbf{k}} & -\overline{V} \\
0 & 0 & -\overline{V}^{\ast} & \varepsilon _{\mathbf{k}\alpha }^{f}%
\end{array}%
\right] ,  \label{HD2}
\end{equation}%
where $\overline{V}=2V\sqrt{D_{\sigma}}[(1-i)\mathbf{k_{x}}+(1+i)\mathbf{k_{y}}]$  and whose spectrum can be written in a Dirac form
\begin{equation}
E_{\pm }(k)=\varepsilon \pm \sqrt{A^{2}(k_{x}^{2}+k_{y}^{2})+M^{2}} ,
\label{Dirac}
\end{equation}%
with $\varepsilon =(\varepsilon _{\mathbf{k}}+\varepsilon _{\mathbf{k}\alpha
}^{f})/2$, $M=(\varepsilon _{\mathbf{k}}-\varepsilon _{\mathbf{k}\alpha
}^{f})/2$ and $A^{2}=8V^{2}D_{\sigma}$. This discussion shows that the X-boson captures the
behavior of the Dirac cones in the spectral density at around the $X$ point of the Brillouin zone as  indicated in Fig. \ref{fig12}.

The correlations appear in the X-boson approach through the renormalization
of the $f$ localized electron energy $\varepsilon_{\mathbf{k}\alpha}^{f}$
and through the quantity $D_{\alpha}=R+n_{f\alpha}$, with $R=\left\langle
X_{0,0}\right\rangle $ and $n_{f\alpha}=\left\langle X_{\alpha
\alpha}\right\rangle $. The quantity ${D}_{\alpha}$ must be calculated
self-consistently through the minimization of the corresponding
thermodynamic potential with respect to the parameter $R$ and the result for
the X-boson parameter $\Lambda$ is
\begin{equation}
\Lambda=\frac{-1}{4\pi^{2}}\int_{-\pi}^{\pi} \int_{-\pi}^{\pi} dk_{x}dk_{y}%
\Biggl\{\overline{E}_{\mathbf{k\alpha}} {\mathcal{F}}^{+}_{\mathbf{k}\sigma,\alpha} + \newline
\nonumber
\end{equation}
\begin{equation}
\frac{\left[2 V^{2}\Delta^{2}{(\mathbf{k})+\overline{E}_{\mathbf{k\alpha}%
}(\varepsilon_{\mathbf{k}\alpha}^{f}- \varepsilon_{\mathbf{k},\sigma}})%
\right]{\mathcal{F}}^{-}_{\mathbf{k}\sigma,\alpha}  }{\sqrt{\left( \varepsilon_{\mathbf{k}}-\varepsilon^{f}_{\mathbf{{k%
}\alpha}}\right)^{2}+4V^{2}D_{\alpha} {\ \Delta}^{2}(\mathbf{k})}}\Biggr\} ,
\label{Eq.36}
\end{equation}
where ${\mathcal{F}}^{+}_{\mathbf{k}\sigma,\alpha}=n_{F}(\omega^{(+)}_{\mathbf{k}%
})+n_{F}(\omega^{(-)}_{\mathbf{k}})$ and 
${\mathcal{F}}^{-}_{\mathbf{k}\sigma,\alpha}=n_{F}(\omega^{(+)}_{\mathbf{k}%
})-n_{F}(\omega^{(-)}_{\mathbf{k}})$.

After the numerical calculation of the parameter $\Lambda$, we calculate the localized $(n_{f})$ and the conduction $(n_{c})$ occupation numbers, employing the mean field Green's functions, Eqs. \ref{Eq3.155}-\ref{Eq3.177}
\begin{eqnarray}
&n_{f,c}=\left( \frac{-1}{\pi }\right) Im\int\limits_{-\infty }^{\infty }d\omega  n_{F}(\omega ) \times \nonumber \\
&\left(\frac{1}{4\pi^{2}}\right)\int_{-\pi}^{\pi} \int_{-\pi}^{\pi} dk_{x}dk_{y}
G_{\mathbf{k}{\sigma },\alpha}^{f,c}(\omega) .  \label{Occupa}
\end{eqnarray}

With these results we calculate the total number of particles $N_{t}=2(n_{f}+n_{c})$, which is maintained constant during all the self-consistent calculations, whereas the chemical potential $\mu$ varies freely.  All the calculations are repeated until the convergence of the short-range antiferromagnetic correlation (SRAFC) function  
$\left< X^{\dagger}_{i\protect\alpha}X_{j\protect\alpha} \right>$ and the X-boson parameters 
$\Lambda$ and  $R$ is attained. 

The density of states is obtained numerically through the relation
\begin{equation}
\rho (\omega )=\left( \frac{-1}{4\pi^{3} }\right) Im \int_{-\pi }^{\pi
}dk_{x}\int_{-\pi }^{\pi }\frac{dk_{y}}{\omega -\omega _{\mathbf{k}}(\pm
)+i\eta },  \label{Density}
\end{equation}%
where $\omega _{\mathbf{k}}(\pm )$ is given by Eq. \ref{Eqn.32}.

To quantify the SRAFC between neighboring f-sites of the lattice we employ the nearest-neighbor
Green's function

\begin{equation}
G_{\mathbf{k}{\sigma },\alpha }^{ij}(z_{n})=\frac{1}{N_s} \sum_{\textbf{k}} 
G_{\mathbf{k}{\sigma },\alpha }^{ff}(z_{n}) e^{i\textbf{k} \cdot (\vec{r_{i}}-\vec{r_{j}})} \label{GNN}
\end{equation}
and considering the tight binding square lattice the SRAFC can be written as \cite{BenHur2000}
\begin{eqnarray}
&\langle X_{i\alpha }^{\dagger }X_{j\alpha }\rangle =\left( \frac{%
-1}{4\pi^{3} }\right) Im\int\limits_{-\infty }^{\infty }d\omega  n_{F}(\omega) \times \nonumber \\
&\int_{-\pi}^{\pi} \int_{-\pi}^{\pi} dk_{x}dk_{y}
\varepsilon_{\mathbf{k}}G_{\mathbf{k}{\sigma },\alpha }^{ff}(\omega).  \label{Corr1}
\end{eqnarray}

In the same way, we can also calculate the Kondo correlation function
\begin{eqnarray}
&\langle X_{i\alpha }^{\dagger }c_{\mathbf{k}\sigma }\rangle=\left( \frac{-1}{4\pi^{3} }\right) Im\int\limits_{-\infty }^{\infty } d\omega  n_{F}(\omega ) \times \nonumber \\
&\int_{-\pi}^{\pi} \int_{-\pi}^{\pi} dk_{x}dk_{y}
G_{\mathbf{k}{\sigma },\alpha}^{fc}(\omega),  \label{Corr2}
\end{eqnarray}%
where  $G_{\mathbf{k}{\sigma },\alpha }^{ff}(z_{n})$ is the localized Green's function
and the $G_{\mathbf{k}{\sigma },\alpha }^{fc}(z_{n})$ is the crossed Green's
function given by Eqs. \ref{Eq3.155} and \ref{Eq3.177}, respectively.

\section{Results and discussion}
\label{sec4}

In this paper we consider the hybridization as $V=0.5t$ and  all the calculations were performed considering  the total number of particles to be constant. We employ $N_{t}=2.0$ for the slave boson, which corresponds to the half-filling case and always produces an insulator independent of the 
$E_{f}$ value. In the X-boson approach, we employ $N_{t}=1.666$, which also corresponds to the half-filling case at the limit of infinite correlation. The $N_{t}$ assumes this value because the strong correlation  shrinks the $f$-band area and the system  can exhibit metallic or TKI phases, depending on the $E_{f}$ value. All the parameters employed in a particular calculation are presented in the corresponding figure. 

\begin{figure}[tbh]
\begin{center}
\includegraphics[clip,width=0.45\textwidth,angle=0.0]
{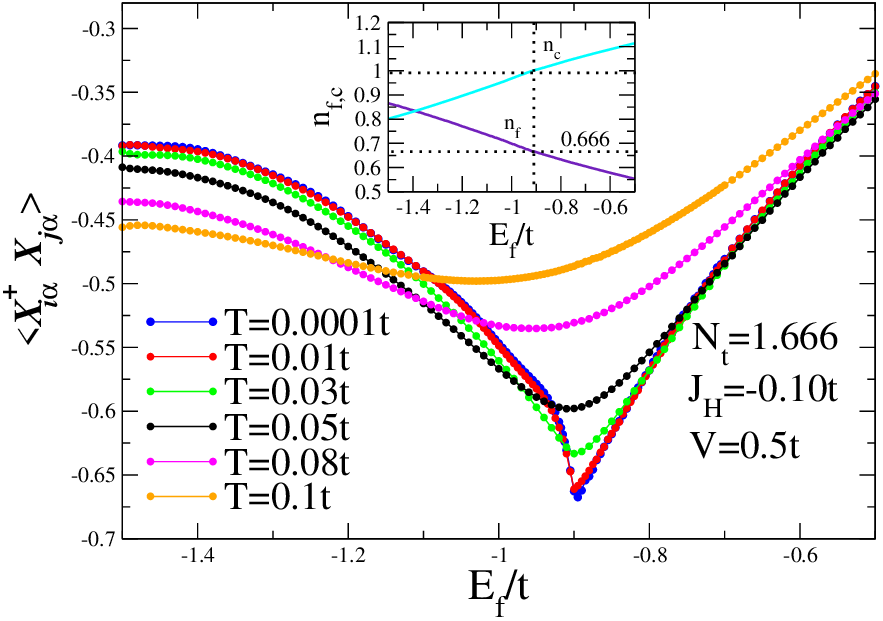}
\end{center}
\caption{(Color online) The  X-boson short-range antiferromagnetic correlation (SRAFC) function  
$\langle X^{\dagger}_{i\protect\alpha}X_{j\protect\alpha} \rangle$   vs. the $E_f$ level
position, for different temperatures values. In the inset we represent the occupation numbers 
$n_{f,c}$.}
\label{fig1}
\end{figure}

\begin{figure}[tbh]
\begin{center}
\includegraphics[clip,width=0.45\textwidth,angle=0.0]
{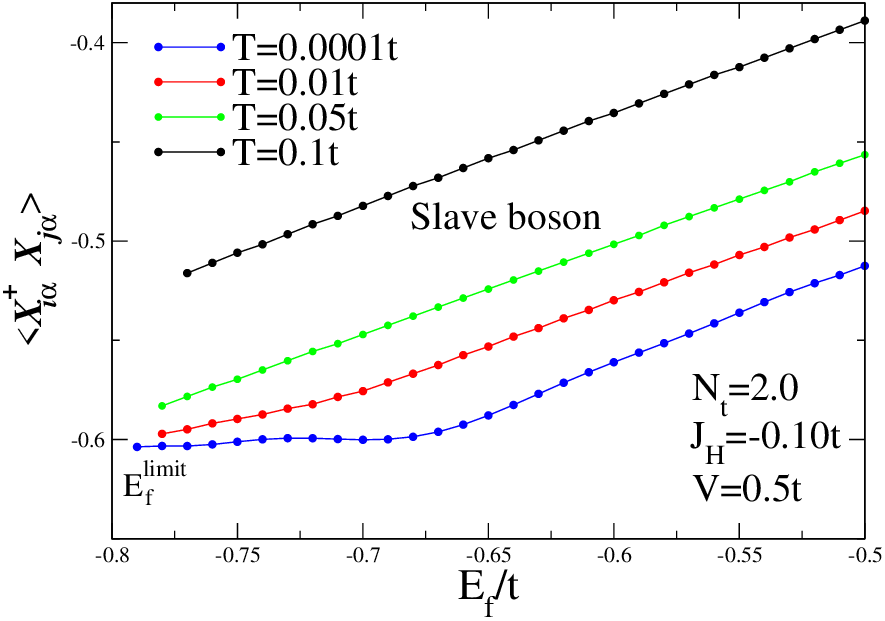}
\end{center}
\caption{(Color online) The SBMFT short-range antiferromagnetic correlation (SRAFC) function 
$\langle X^{\dagger}_{i\protect\alpha}X_{j\protect\alpha} \rangle$   vs. the $E_f$ level
position, for different temperatures values.}
\label{fig2}
\end{figure}

In Figs. \ref{fig1} and \ref{fig2}, we plot the SRAFC function for the
X-boson and SBMFT, respectively, both as a function of the $f$ level position
$E_{f}$, for different temperatures. At high 
temperatures, in both cases, the functions are smooth, indicating that the
magnetic moments of the atoms are uncorrelated and distributed at random. As the temperature
is lowered, the SRAFC becomes more effective and the X-boson correlation functions, plotted in  
Fig. \ref{fig1}, develop a sharp minimum in the region of the formation of the TKI. The  topological Kondo insulator  is obtained when $n_{f}=0.666$, and  as the $c$-band is uncorrelated, it continues to have $n_{c}=1.0$, as indicated in the inset of the Fig.\ref{fig1}. On the other hand, Fig. \ref{fig2} shows that the SBMFT is not able to capture these antiferromagnetic  correlations properly; it only develops  a plateau when $n_{f}$ goes to the unit and $E_{f}$ goes to $E^{limit}_{f}$, as indicated on the left corner of the figure, where SBMFT breaks down. The SBMFT exhibits insulator behavior for any $E_{f}$ value, while the X-boson exhibits topological insulator behavior only at around the $E_{f}$ value, where the correlation function 
$\langle X^{\dagger}_{i\alpha} X_{j\alpha}\rangle$ exhibits its minimum  and metallic behavior for other values of $E_{f}$.

\begin{figure}[tbh]
\begin{center}
\includegraphics[clip,width=0.45\textwidth,angle=0.0]
{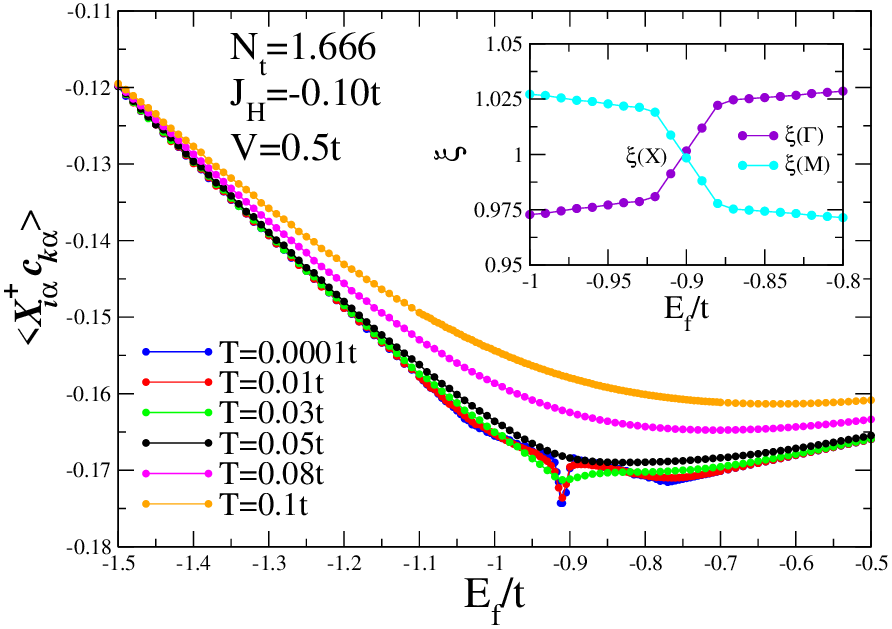}
\end{center}
\caption{(Color online) The X-boson Kondo correlation function for different
temperatures values, $\langle X^{\dagger}_{i\protect\alpha} c_{\mathbf{k}\protect\sigma}\rangle$ vs. the $E_f$ level position. In the inset we represent the $\xi$ closing gap equation corresponding to the high symmetry points of the Brillouin zone: $\Gamma$, $X$ and 
$M$.}
\label{fig3}
\end{figure}

In Fig. \ref{fig3},  we plot the Kondo correlation function 
$\langle X^{\dagger}_{i\alpha} c_{\mathbf{k}\sigma}\rangle$ for the X-boson  
as a function of the $E_f$ level position, for different 
temperatures and for the same parameter values employed in Fig. \ref{fig1}.  At high temperatures, the functions are smooth, indicating that the magnetic moments of the atoms are uncorrelated and distributed at random. As the temperature is lowered, the 
$\langle X^{\dagger}_{i\alpha} c_{\mathbf{k}\sigma}\rangle$ develops a sharp  minimum in the region where the SRAFC is more intense,  indicating that both processes, the SRAFC and Kondo correlations,  act in the development of the topological Kondo gap (TKG) formation. In the $2D$ case studied here, we obtain a single-band inversion only at the $X$ point of the Brillouin zone, which agrees with earlier studies \cite{Takimoto11} and from the analytical expression  \cite{Sigrist14}
\begin{eqnarray}
&\xi=-D_{\sigma} J_{H} \langle X^{\dagger}_{i\alpha} X_{j\alpha}\rangle/2= \nonumber \\
&1+\tilde{\varepsilon}_{f}/2t[cos(k_{x})+cos(k_{y})] ,
\label{Close}
\end{eqnarray}
with $\tilde{\varepsilon}_{f}=E_{f}+\Lambda-\mu$,  we obtain the closure of the gap at the  $X$ point, when $\xi(X) \rightarrow 1$, as indicated in the inset of Fig. \ref{fig3} and \ref{fig12}.  For the particular set of parameters employed in our calculations, this limit is attained for $E_{f}=-0.912t$,  when  $\tilde{\varepsilon}_{f} \rightarrow 0$ and $\mu=0$.

The interesting point of the earlier  results is that the X-boson approach is able to capture 
the competition between SRAFC and the Kondo effect in the IV region, characterized by the development of a sharp minimum in the correlation functions at  low temperatures. SRAFC favors the formation of magnetic moments on the atoms, and at the same time the existence of those moments opens  up the possibility of spin-flip scattering by the conduction electrons, generating the Kondo effect. Since we are in the intermediate valence region, with the localized occupation number being $%
n_{f}=0.666$, none of these correlation processes wins over the other, and they
act in a cooperative way to open the TKG. The
competition between these two processes at the Kondo limit, where $n_{f}
\simeq 1$, is well described by the Doniach diagram \cite{Doniach1977,Coqblin97}. At
this limit, this competition is stronger, and the system attains some
magnetic order, generally antiferromagnetic, or goes to the heavy fermion
Kondo regime, where the system is a Fermi liquid and never develops magnetic
order, even at very low temperatures. 

\begin{figure}[tbh]
\begin{center}
\includegraphics[clip,width=0.45\textwidth,angle=0.0]
{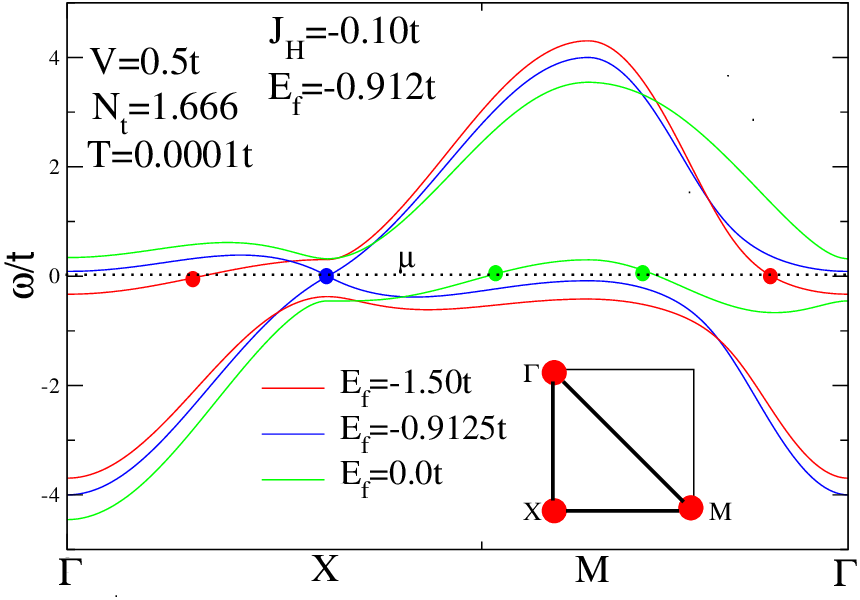}
\end{center}
\caption{(Color online) Energy bands governed by the X-boson approach. The calculations were performed along the high symmetry points of the Brillouin zone 
$\Gamma \rightarrow X  \rightarrow M \rightarrow \Gamma$.}
\label{fig4}
\end{figure}
In Fig. \ref{fig4}, we represent the X-boson energy spectrum for some representative $E_{f}$ values:  $E_{f}=0.0t$ in the low occupation region, $E_{f}=-0.912t$ at the critical point,  where the condition for the development of the TKI occurs, and  at $E_{f}=-1.5t$ in the Kondo region. The calculations were performed along the high symmetry points of the Brillouin 
zone: $\Gamma \rightarrow X  \rightarrow M \rightarrow \Gamma$.   In the inset of the figure, we show the exact position where the gap closes with the formation of the Dirac cone at the $X$ point, where a band inversion occurs. The presence of correlations changes the situation in relation to the previous analysis employing uncorrelated bands \cite{Sigrist14} or SBMFT \cite{Dzero13,Dzero12}, where the topological transition occurs between insulators states. Here the transition occurs between metallic states, represented by the energies $E_{f}=0.0t$ and  $E_{f}=-1.5t$, where the chemical potential crosses band states twice, which  indicates a metallic  topologically trivial phase. The system also exhibits a topologically non-trivial phase, which defines the TKI,  where a band inversion occurs and  the chemical potential crosses band states only once. This situation is represented in the phase diagram of Fig. \ref{fig8}.

\begin{figure}[tbh]
\begin{center}
\includegraphics[clip,width=0.45\textwidth,angle=0.0]
{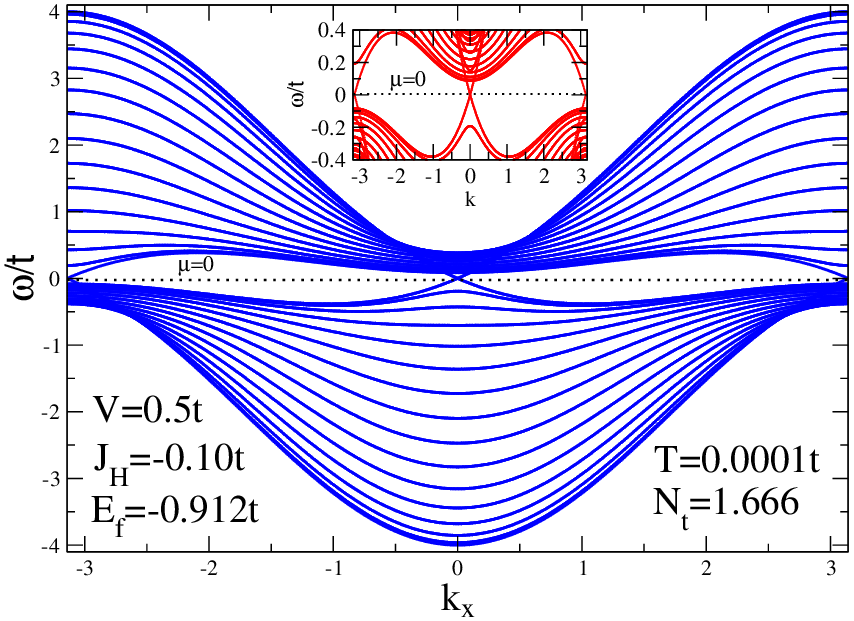}
\end{center}
\caption{(Color online) Energy spectrum governed by the X-boson approach
along the $x-$axis, $k=k_{x}$, with open boundaries in the $y$ direction.}
\label{fig5}
\end{figure}
In Fig. \ref{fig5}, we represent the X-boson energy spectrum at the critical point 
$E_{f}=-0.912t$, where the condition for the
development of the TKI occurs. The calculations were performed along the 
$x$-axis, $k=k_{x}$ with open boundaries in the $y$ direction.  The figure shows the formation of Dirac cones at  $X$ points corresponding to the critical  $ E_{f}=-0.912t$ value and the 
chemical potential $\mu=0$. The band structure defines a strong topological insulator \cite{Dzero12}. In the inset of the figure we show the exact positions where the gap closes. 
\begin{figure}[tbh]
\begin{center}
\includegraphics[clip,width=0.45\textwidth,angle=0.0]
{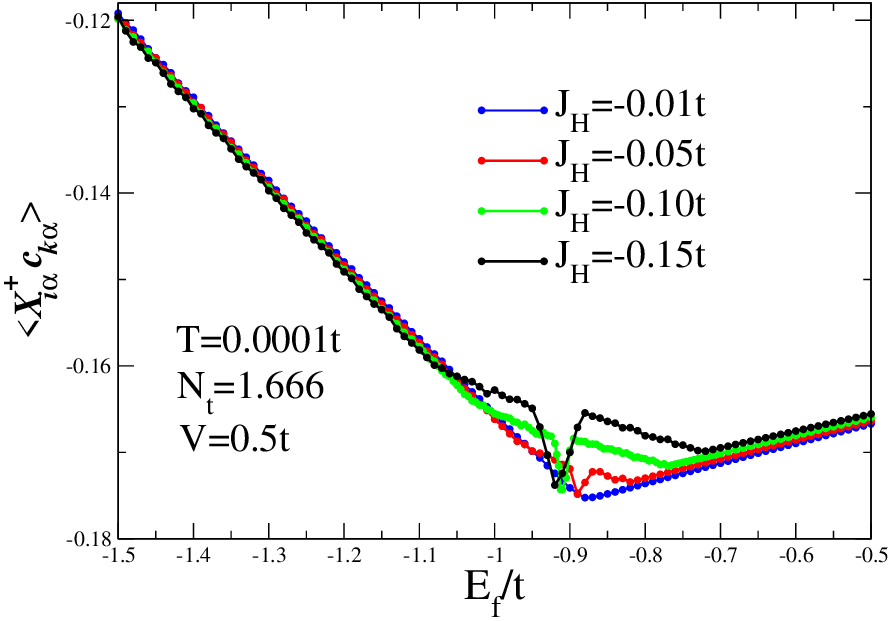}
\end{center}
\caption{(Color online) The Kondo correlation function  
{$\langle X^{\dagger}_{i\protect\alpha} c_{\mathbf{k}\protect\sigma}\rangle$}, for different 
$\protect J_{H}$ values, vs. $E_f$ level position.}
\label{fig6}
\end{figure}

In Fig. \ref{fig6}, we plot the Kondo correlation function 
$\langle X^{\dagger}_{i\alpha} c_{\mathbf{k}\sigma}\rangle$ for the X-boson, considering different 
$J_{H}$ values, as a function of the $E_f$ level position. As we
increase the SRAFC, characterized by the parameter $J_{H}$, the range of values around the minimum increases and the minimum of the curve
points directly to the position of the $E_{f}$ value that defines the TKI. The slight displacement of the $E_{f}$ values  to the left indicates the increase of the Kondo correlations present in the system once the localized occupation numbers increases, as indicated in the inset of Fig. \ref{fig1}.

\begin{figure}[tbh]
\begin{center}
\includegraphics[clip,width=0.40\textwidth,angle=0.0]
{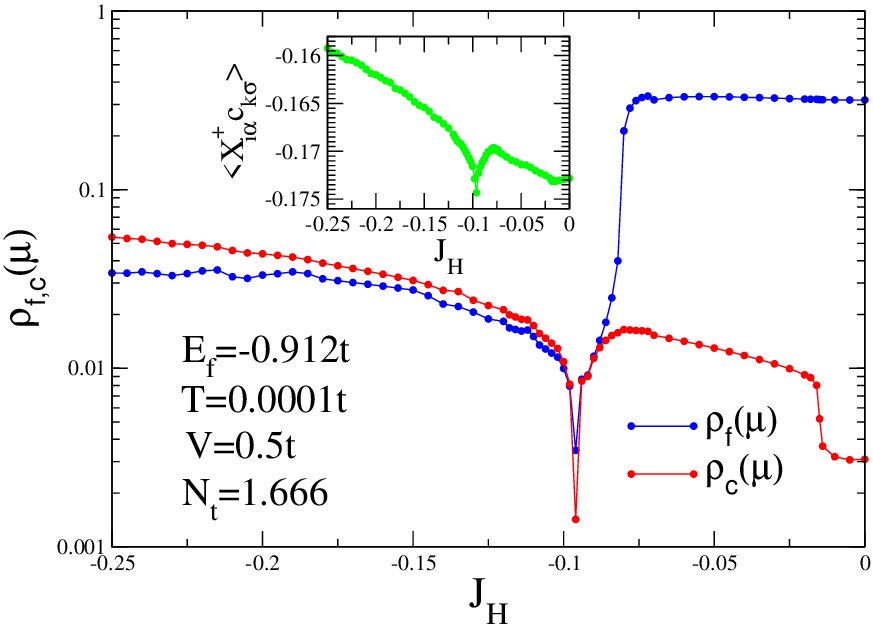}
\end{center}
\caption{(Color online) X-boson density of states of the localized 
and conduction electrons at the chemical potential $\mu=0$, as a function of $J_{H}$. In the inset, we
represent the Kondo correlation function in  logarithmic scale.}
\label{fig7}
\end{figure}
In Fig. \ref{fig7}, we represent the X-boson density of states of the localized  and conduction electrons at the chemical potential $\mu=0$, as a function of $J_{H}$. This result indicates that the TKG is induced by the SRAFC. As we vary the $J_{H}$ parameter, at around 
$J_{H} \simeq 0.10t$ the curve exhibits a sharp minimum.  In the inset of the figure,  we plot the Kondo correlation function, which exhibits a minimum in exactly the same region where the TKG appears in the main panel, which indicates a close relation between the Kondo effect and SRAFC in the formation of the TKG. 

\begin{figure}[tbh]
\begin{center}
\includegraphics[clip,width=0.45\textwidth,angle=0.0]
{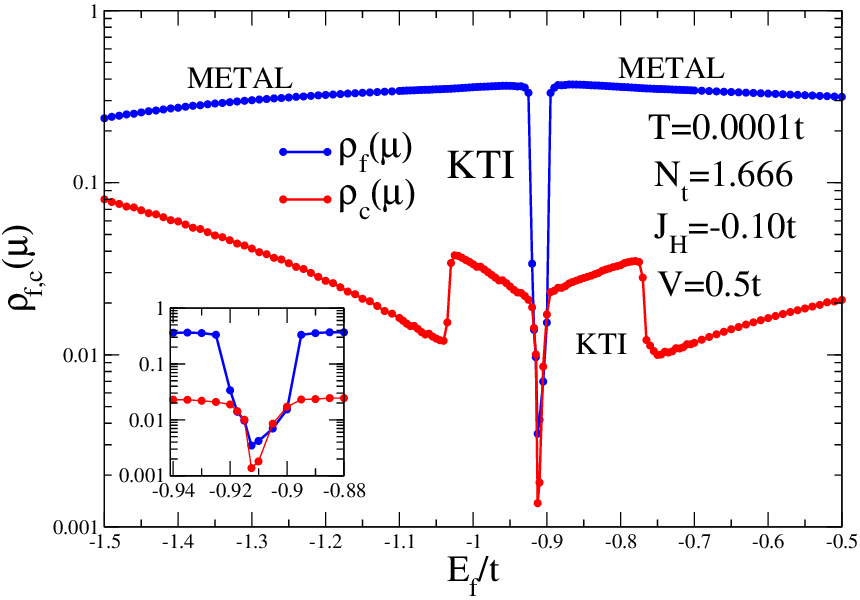}
\end{center}
\caption{(Color online) X-boson density of states of localized $\protect\rho_{f}(%
\protect\mu)$ and conduction $\protect\rho_{c}(\protect\mu)$ electrons, at
the chemical potential $\protect\mu$, as a function of the localized level $%
E_{f}$. The capital letters in the figure represent: metallic region (Metal) and topological Kondo insulator (TKI).}
\label{fig8}
\end{figure}

\begin{figure}[tbh]
\begin{center}
\includegraphics[clip,width=0.45\textwidth,angle=0.0]{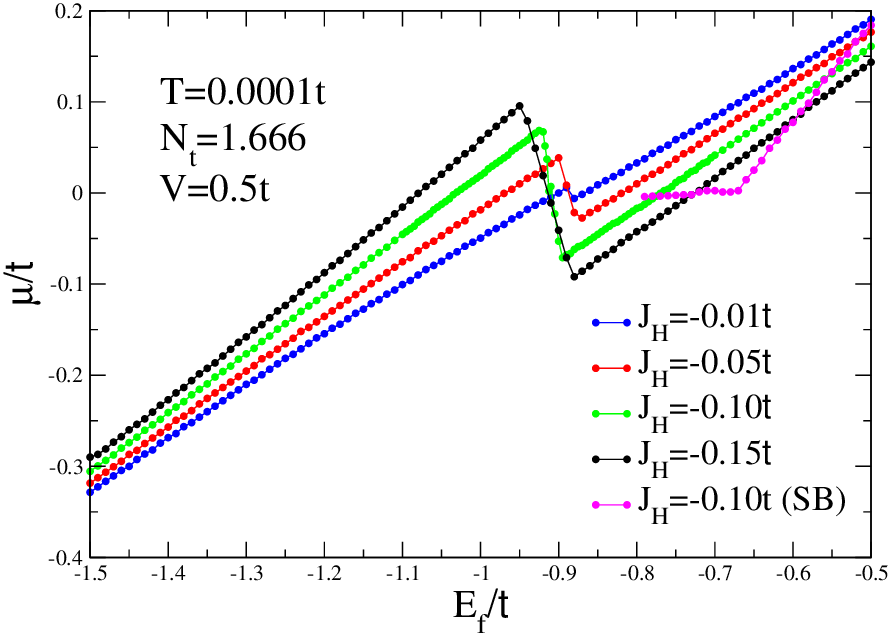}
\end{center}
\caption{(Color online) Chemical potential as a function of the localized
level energy $E_{f}$ calculated employing the X-boson approach and the
SBMFT, for different $J_{H}$ values.}
\label{fig9}
\end{figure}
In Fig. \ref{fig8}, we plot the X-boson density of states of localized $%
\rho_{f}(\mu)$ and conduction $\rho_{c}(\mu)$ electrons, at the chemical
potential $\mu$, as a function of the localized level $E_{f}$.  The system exhibits a metallic behavior for all the $E_{f}$ values except in the   topological Kondo insulator 
(TKI) region \protect{$-0.925t \lesssim E_{f} \lesssim -0.895t$}.  As discussed in the beginning of this section, only when $N_{t}=1.666$ does the development of the topological Kondo insulator occur, and the chemical potential is located  inside the gap. This result contrast with the SBMFT \cite{Dzero12,Tran12}, where for total occupation $N_{t}=2.0$, and  independent of the $E_{f}$ values the system is always an insulator, and the magnitude of the gap increases indefinitely as $E_{f}$
goes to the Kondo limit.  Our phase diagram is consistent with pressure experiments of $SmB_{6}$ which show an insulator-metal transition due to the gap closing, followed by a long range magnetic order at approximately $P=10 \hspace{1.0mm} GPa$ \cite{Derr08}. In the inset of the figure, we plot a detail of the TKG; the minimum of $\rho_{f,c}(\mu)$ occurs at $E_{f}=-0.912t$. 

In Fig. \ref{fig9}, we plot the X-boson chemical potential as a function of the
localized level $E_{f}$ for representative $J_{H}$ values. The most prominent fact here is the sharp
transition exhibited by the chemical potential when crosses the Kondo topological
region. It is worth pointing out that when the magnitude of $J_{H}$ increases, the TKG opens at a more negative $E_{f}$ value position (from $E_{f}=-0.88t$ for $J_{H}=-0.01t$, to $E_{f}=-0.93t$ for 
$J_{H}=-0.15t$), indicating that the SRAFC acts to reinforce the Kondo effect. 

In the X-boson approach, the strong correlation is reflected in the shrinking of the $f$-band area to $D_{\alpha}=0.666$ whereas the conduction band, which is uncorrelated, continues to have an area equal to the unit. For the sake of comparison, we also plot the SBMFT results calculated employing $N_{t}=2.0$ and $J_{H}=-0.10t$. The SBMFT exhibits a plateau at around the region $-0.80t \lesssim E_{f} \lesssim -0.65$, which is the Kondo region of the model ($n_{f}=1.0$ for $E_{f}=-0.80$). For $E_{f} \lesssim -0.80$, the SBMFT breaks down. This result indicates that in the IV region, the SBMFT does not capture the strong electronic correlation of the localized f-electrons exhibited by the X-boson results.

\begin{figure}[tbh]
\begin{center}
\includegraphics[clip,width=0.45\textwidth,angle=0.0]
{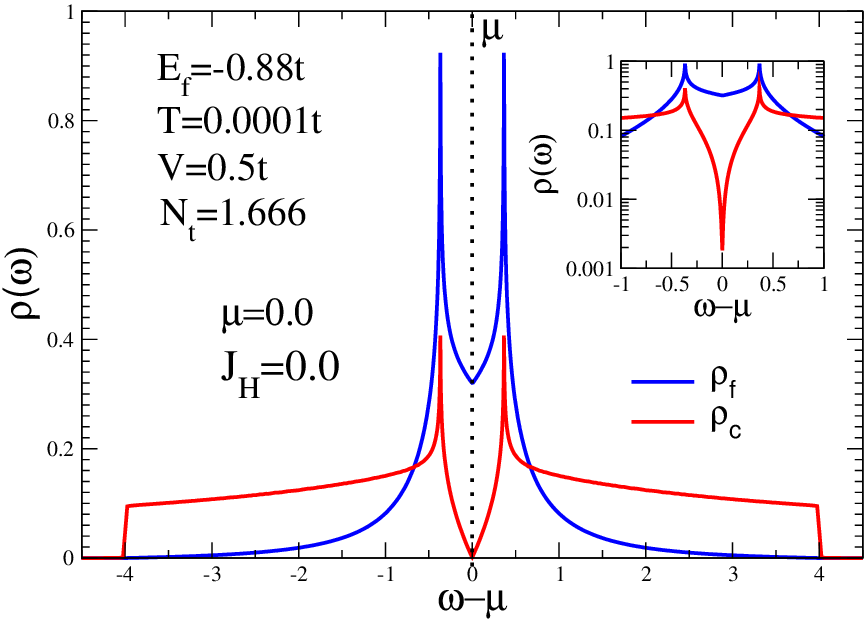}
\end{center}
\caption{(Color online) X-boson density of states of the localized and conduction
electrons with $J_{H}=0$. In the inset we
represent the density of states, in a logarithmic scale, to show the
V-shape character of the conduction topological Kondo gap.}
\label{fig10}
\end{figure}
In Fig. \ref{fig10}, we represent the X-boson density of states of the localized and
conduction electrons, without considering the SRAFC ($J_{H}=0.0$). The particular $\mathbf{k}$ dependence of the hybridization (cf. Eq. \ref{Hyb}), characterized by the
representation $\Gamma^{(1)}_{8}$, leads to the opening of a V-shaped
conduction electrons hybridization gap, as indicated in the inset of the
figure. However, the localized electron density of states remains finite within
all the frequency range, resulting in a metallic phase. 

\begin{figure}[tbh]
\begin{center}
\includegraphics[clip,width=0.45\textwidth,angle=0]
{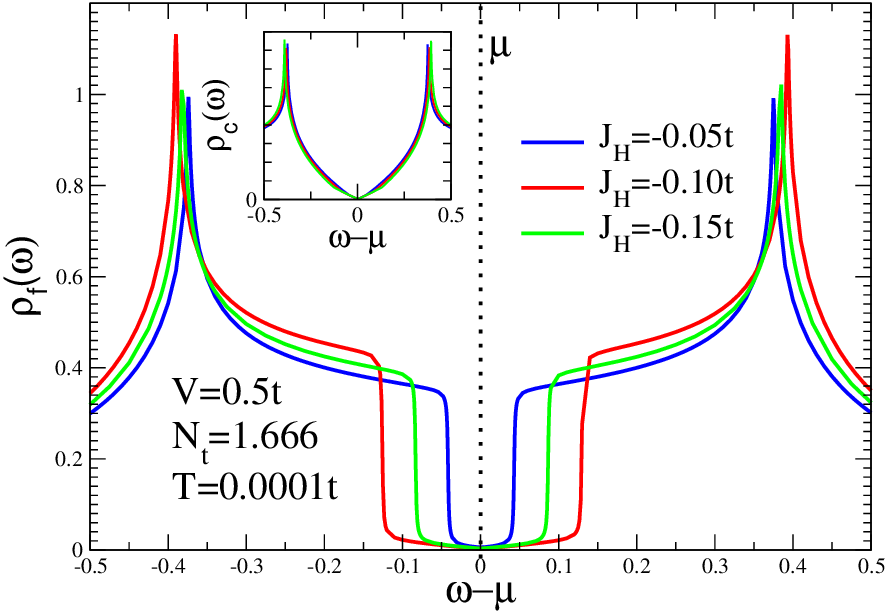}
\end{center}
\caption{(Color online) X-boson density of states of the localized and conduction
electrons considering different $J_{H}$ values.}
\label{fig11}
\end{figure}
In Fig. \ref{fig11}, we plot the X-boson density of states of the localized and
conduction electrons, but now turning on the SRAFC. We
consider different $J_{H}$ and  $E_{f}$ values: $(J_{H}=-0.05t \hspace{0.1cm};\hspace{0.1cm} E_{f}=-0.896t)$, 
$(J_{H}=-0.10t \hspace{0.1cm};\hspace{0.1cm} E_{f}=-0.912t)$, and $(J_{H}=-0.15t \hspace{0.1cm};\hspace{0.1cm} E_{f}=-0.93t)$. Now, due to the existence of the SRAFC,  the localized  density of states opens a gap at the chemical potential (TKG). The curves show an enlargement of the gap as the magnitude of the  
$J_{H}$ parameter increases  as well as a steep increase of the edge peaks.  At the same time, the localized  $E_{f}$ values vary from  $E_{f}=-0.896$ to $E_{f}=-0.93t$, which indicates an increase of the Kondo character of the system. In the inset, we represent the conduction density of states, which is practically insensitive to the SRAFC. From this result it
is clear that the $\Gamma^{(1)}_{8}$ representation by itself is not
sufficient to describe the physics of the topological Kondo insulators like
the $SmB_{6}$, which exhibits a true gap in the density of states, but it can
be relevant to the study of the Kondo insulator $CeNiSn$, which exhibits a
spin gap originated from SRAFC \cite{Park98,Sato05} and a V-shaped density
of states \cite{Ikeda96,Nakamura96,JuanaMoreno2000}.

\begin{figure}[tbh]
\begin{center}
\includegraphics[clip,width=0.45\textwidth,angle=0]
{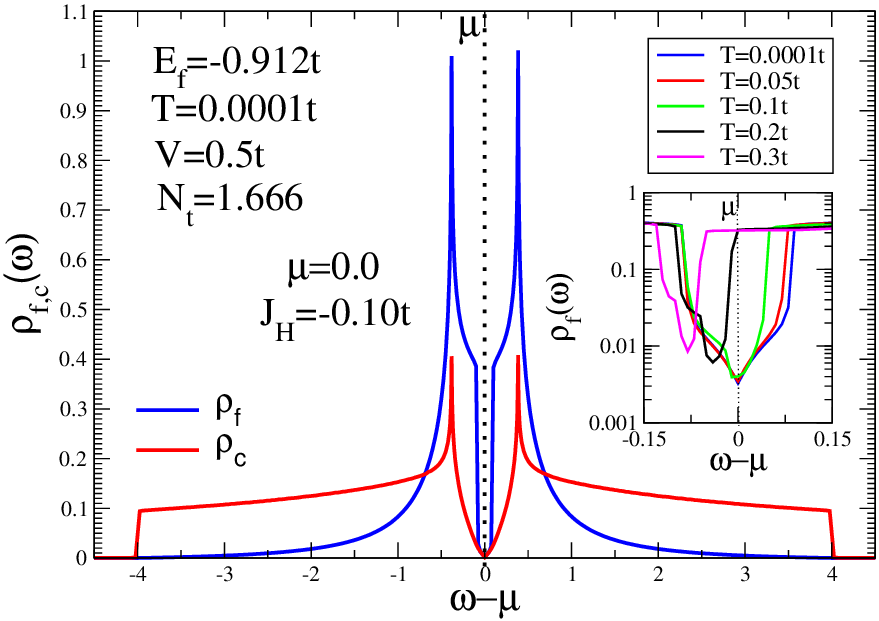}
\end{center}
\caption{(Color online) X-boson density of states of the localized and conduction
electrons. In the inset we
represent the density of states in a logarithmic scale for different  temperatures.}
\label{fig12}
\end{figure}

In Fig. \ref{fig12}, we plot the X-boson density of states of the localized and
conduction electrons as a temperature function. Now, due to the existence of SRAFC, both the localized and the conduction density of states open a gap at the chemical potential (TKG), 
producing a V-shaped gap at around the critical $E_{f}=-0.912t$ value. In the inset
of the figure, we plot a temperature dependence of this V-shaped gap. As we increase the temperature from $T=0.0001t$ to $T=0.01t$ the gap remains unchanged. However, for temperatures $T>0.05t$, the thermal effects act in a more effective way, and the gap is reduced and undergoes a displacement to below the chemical potential and disappears at high temperatures. At $T=0.2t$, the localized density of states crosses the chemical potential, and the system undergoes an insulator-metal transition. As happens in real Kondo insulators, once at high temperatures they behave as dirty metals.

\section{Conclusions}
\label{sec5}

Considering several experimental results, obtained via inelastic neutron
scattering (INS),   in the Kondo insulators $YbB_{12}$ \cite{Mignot05}, $SmB_6$ \cite{Alekseev09} and $CeNiSn$ \cite{Park98,Sato05}, we employed the X-boson method 
\cite{Xboson} to take into account the short-range antiferromagnetic (SRAFC) correlations in these systems.

We have shown that in the intermediate valence region
(IV regime), the SRAFC favors the formation of magnetic moments on the atoms, and
at the same time, the existence of such moments open up the possibility of
spin-flip scattering by the conduction electrons, generating the Kondo
effect. Contrary to the heavy fermion limit, described by the Doniach
diagram \cite{Doniach1977}, whose correlation effects generate a strong
competition between magnetic moments and conduction electron scattering, inducing the system 
to attain some magnetic order or remain in
the Kondo Fermi liquid regime,  in the intermediate valence region those correlations act in a cooperative way to open the spin gap and generate the TKG. We also study the evolution of the TKG with increasing temperature, and we obtain that at high temperatures the system undergoes an insulator-metal transition, as happens in real Kondo insulators.

We also calculated the band structure along the high symmetry points of the Brillouin zone  $\Gamma \rightarrow X  \rightarrow M \rightarrow \Gamma$. We showed that the gap closes  at the $X$ point, with the formation of the Dirac cone with the corresponding  band inversion. This result agrees with an earlier study \cite{Takimoto11}. The striking point here is that the presence of strong correlations  completely change the situation in relation to previous analysis employing uncorrelated bands \cite{Sigrist14} or SBMFT \cite{Dzero13,Dzero12}, where the topological transition occurs between insulator states. Here the topological transition occurs between metallic states, which shows that the X-boson method captures the Dirac cone structure of the TKI.  

We calculated the Kondo and the SRAFC correlation functions, showing that the range of values at around the minimum of these functions  grow  with an  increase in the parameter $J_{H}$, which controls the strength of the SRAFC. We showed that the position of the minimum of these functions defines the range within which the TKI appears in the phase diagram. We also calculated the phase diagram of Fig. \ref{fig8}, which shows that as we vary the $E_{f}$ level position, from the empty lattice  to the Kondo regime, the system develops two phases: metallic and TKI, where this last occurs in a very restricted region of $E_{f}$ values, and is formed due to the existence of SRAFC. This result is consistent with pressure experiments of $SmB_{6}$, which show an insulator-metal transition due to the closing of the TKG followed by a magnetic long range order at approximately 
$P=10 \hspace{1.0mm} GPa$ \cite{Derr08}

We also presented localized and conduction density of states curves, which show the opening of a V-shaped TKG at the chemical potential, due to the presence of SRAFC. This kind of result is a consequence of the use of the $\Gamma^{(1)}_{8}$ representation, which by itself is not sufficient to describe the physics of the topological Kondo insulators like  $SmB_{6}$,  which exhibit a true gap in the density of states. To obtain a more realistic description of the TKI, it is necessary to include the other $\Gamma^{(2)}_{8}$ representation  in the formalism, as well as to consider the problem in three dimensions.  However, the results obtained here  can be relevant to the study of the Kondo insulator 
$CeNiSn$, which exhibits a spin gap originated from SRAFC \cite{Park98,Sato05} and a V-shaped density
of states \cite{Ikeda96,Nakamura96,JuanaMoreno2000}.

\ack
We are thankful for the financial support of the Brazilian National Research Council (CNPq), DIEB (Colombia National University), and the Colombian National Science Agency COLCIENCIAS (Grant FP44842-027-2015). R. Franco and J. Silva-Valencia are grateful to ICTP - Trieste. 

\medskip
\appendix

\section{Heisenberg mean field}
\label{sec6}

In this Appendix we transform the Heisenberg Hamiltonian into a one-particle term by employing a mean field approximation,
\begin{equation}
\mathcal{H}_{H}=-J_{H}\sum\limits_{\left\langle i,j\right\rangle
}\mathbf{S}_{i} \cdot \mathbf{S}_{j} ,
\label{E1}
\end{equation}
where $i,j \in \{1, . . . ,N\}$ label  lattice sites and  $\left\langle i,j\right\rangle$
denote pairs of nearest-neighbors (NN) sites.

\begin{eqnarray}
&\mathcal{H}_{H}=-J_{H}\sum\limits_{\left\langle i,j\right\rangle} 
S_{z}\left(i\right) S_{z}\left(j\right) - \nonumber \\
&\frac{1}{2} J_{H}\sum\limits_{\left\langle i,j\right\rangle}\left[S_{+}\left(  i\right)S_{-}\left(  j\right)+S_{-}\left(  i\right)  S_{+}\left(  j\right)\right]  .
\label{E11}
\end{eqnarray}
As in this work we will be restricted to the nonmagnetic phase, the first term of Eq. \ref{E11} does not contribute and we only  consider the second one. We consider only two states, with pseudospin label $\alpha=\pm$, belonging to the representation $\Gamma_{8}^{(1)}$ of the multiplet state at the site $j$.

Introducing the Hubbard operators  $X_{a,b}(j)$ in a site $j$: 
$X_{+,-}\left\vert -\right\rangle=\left\vert +\right\rangle$; 
$X_{-,+}\left\vert +\right\rangle=\left\vert -\right\rangle$;
$X_{+,+}\left\vert +\right\rangle=\left\vert +\right\rangle$;
$X_{-,-}\left\vert -\right\rangle=\left\vert -\right\rangle$,
we can write Eq. (\ref{E11}) as
\begin{eqnarray}
&\mathcal{H}_{H}=-J_{H}\sum\limits_{\left\langle i,j\right\rangle}
[X_{+,-}\left(i\right) X_{-,+}\left(j\right)+ \nonumber \\
&X_{+,-}\left( j\right)X_{-,+}\left(i\right)], \label{E3}
\end{eqnarray}
and employing the Hubbard operators relations
\begin{equation}
X_{+,-}\left(  i\right)=X_{+,0}\left(  i\right)  X_{0,-}\left(  i\right) ,
\label{H1}
\end{equation}
\begin{equation}
X_{-,+}\left(j\right)=X_{-,0}\left(j\right)  X_{0,+}\left(  j\right) ,
\label{H2}
\end{equation}
we can introduce the following mean field approximation
\begin{eqnarray}
& 2\left[ X_{+,-}\left(  i\right)  X_{-,+}\left(  j\right)  +X_{+,-}\left(  j\right)
X_{-,+}\left(  i\right) \right] =\nonumber\\
& X_{+,0}\left(  i\right)  X_{0,+}\left(  j\right)
\left\langle X_{0,-}\left(  i\right)  X_{-,0}\left(  j\right)  \right\rangle + \nonumber \\
& X_{+,0}\left(  j\right)  X_{0,+}\left(  i\right)  \left\langle X_{0,-}\left(
j\right)  X_{-,0}\left(  i\right)  \right\rangle + \nonumber\\
&   X_{-,0}\left(  i\right)  X_{0,-}\left(  j\right)  \left\langle
X_{0,+}\left(  i\right)  X_{+,0}\left(  j\right)  \right\rangle + \nonumber \\
& X_{-,0}\left(  j\right)  X_{0,-}\left(  i\right)  \left\langle X_{0,+}\left(
j\right)  X_{+,0}\left(  i\right)  \right\rangle,
\label{E5}
\end{eqnarray}

\noindent and the Hamiltonian can be written in the form
\begin{eqnarray}
&\mathcal{H}_{H}=-\frac{1}{2}J_{H}\sum\limits_{i,j,\alpha} \langle 
X^{\dagger}_{0,\overline{\alpha}} \left(i\right)X_{0,\overline{\alpha}}\left(  j\right)  
\rangle \times \nonumber \\
&\left[ X_{0\alpha}^{\dagger}\left( i\right)X_{0\alpha}\left(  j\right)   \right] ,
\label{E12}
\end{eqnarray}
where we use the additional mean-field approximation
\begin{equation}
\left\langle X_{0,\alpha}\left(  j\right)  X_{\alpha,0}\left(  i\right)
\right\rangle =\left\langle X_{0,\alpha}\left(  i\right)  X_{\alpha,0}\left(
j\right)  \right\rangle , \label{E6}
\end{equation}
and comparing with the effective hopping Hamiltonian
\begin{equation}
H_{f}^{\prime}=\sum_{i,j,\alpha}t_{i,j,\sigma}\ X_{0\alpha}^{\dagger}\left( i\right)X_{0\alpha}\left(  j\right)  ,
\label{E12}%
\end{equation}
we obtain
\begin{equation}
t_{i,j,\alpha}=-\frac{1}{2}J_{H} \langle 
X^{\dagger}_{0,\overline{\alpha}} \left(  i\right)  X_{0,\overline{\alpha}}\left(  j\right) \rangle  . \label{E13}%
\end{equation}

Since we are studying the paramagnetic case, we simplify the notation by writing the  X-boson short-range antiferromagnetic correlation  function as 
$\left< X^{\dagger}_{i\protect\alpha}X_{j\protect\alpha} \right>$ and Eq. \ref{E13} becomes

\begin{equation}
t_{i,j,\alpha}=-\frac{1}{2}J_{H} \langle X^{\dagger}_{i\protect\alpha}X_{j\protect\alpha} \rangle  . \label{E133}%
\end{equation}

\section{Details of the cumulant expansion}
\label{sec7}

Considering the Heisenberg Hamiltonian, given by Eq.\ref{E121}, as an additional perturbation in the main Hamiltonian (Eq.\ref{HGCE}), the cumulant expansion follows the same lines
of the derivation {\cite{Xboson,FFM}} used in the absence of that
perturbation. The only difference in the resulting diagrams is that there
are now two types of edges: besides the hybridization edge appearing in the
early derivation, there is another type of edge associated with the hopping of
the $f$- electrons. As before, one has to construct all the topologically
different diagrams, but there are many new diagrams corresponding to the
presence of the hopping edge.

The simplest calculations are usually performed when one uses imaginary
frequency and reciprocal space, and the transformation from imaginary time
to frequency is done exactly as before {\cite{Xboson,FFM}, but to transform
to reciprocal space it is also necessary to express the hopping constant }$%
t_{i,j}$ employing its Fourier transform $\overline{E}_{\mathbf{k\alpha}}$:
\begin{equation}
t_{i,j}=\frac{1}{N_{s}}\sum_{\mathbf{k}_{i},\mathbf{k}_{j}}(\overline{E}_{%
\mathbf{k}_{i}})\ \delta _{\mathbf{k}_{i},\mathbf{k}_{j}}\exp [\mathbf{k}%
_{i}.\mathbf{R}_{i}]\ \exp [-\mathbf{k}_{j}.\mathbf{R}_{j}],  \label{Eq2.5}
\end{equation}%
and from $t_{i,j}=t_{j,i}^{\ast }$ it follows that $\overline{E}_{\mathbf{k\alpha}}$
is real (this $\overline{E}_{\mathbf{k\alpha}}$ should not be confused with the $%
E_{\mathbf{k,\sigma }}$ of Eq.(\ref{Hconductio})). It is straightforward to
show that in reciprocal space there is conservation of $\mathbf{k}$ along
the hopping edges, and that each of them multiplies the corresponding
diagram's contribution to a factor $\overline{E}_{\mathbf{k\alpha}_{i}}\ \delta
_{\mathbf{k}_{i},\mathbf{k}_{j}}$.

The mean-field chain approximation (CHA) \cite{Xboson,FFM}, gives simple 
but useful approximate propagators, obtained in the
cumulant expansion by taking the infinite sum of all the diagrams that
contain ionic vertices with only two lines. The laborious calculation of the
general treatment is rather simplified in this case, and
the inclusion of the perturbation $H_{f}^{\prime }$ in the calculation is
fairly simple in this approximation.

The only difference in the CHA diagrams when we consider the hopping between
the $f$-electrons is the appearance of any number of consecutive local
vertices, joined by hopping edges that contribute a factor $t_{i,j}$ when
they join the two sites $i$ and $j$, while the contribution of the
corresponding hybridization edges is not altered. The whole calculation
becomes very simple if we regroup the diagrams so that all the possible sums
of contiguous local vertices are considered as a single entity. We can
calculate the partial contribution of the set of all those diagrams that
join a fixed pair of conduction vertices $j$ and $j^{\prime }$ (and do not
contain any conduction vertex inside the collection): {we obtain (with $%
\alpha =(oa)$):
\begin{eqnarray}
&\widetilde{G}_{j\prime \alpha \prime ;j\alpha }^{ff}(z_{n})=\delta _{\alpha
\prime ,\alpha }\ \delta _{j\prime ,j}G_{f,\alpha }^{o}(z_{n})+ \nonumber \\
&G_{f,\alpha}^{o}(z_{n})t_{j^{\prime },j}G_{f,\alpha }^{o}(z_{n})+\nonumber \\
&G_{f,\alpha }^{o}(z_{n})\sum_{j_{1}}t_{j\prime ,j_{1}}G_{f,\alpha
}^{o}(z_{n})t_{j_{1},j}G_{f,\alpha }^{o}(z_{n})+ \nonumber \\
&G_{f,\alpha }^{o}(z_{n})\sum_{j_{1}}t_{j\prime ,j_{1}}G_{f,\alpha
}^{o}(z_{n}) \times \nonumber \\
&\sum_{j_{2}}t_{j_{1},j_{2}}G_{f,\alpha
}^{o}(z_{n})t_{j_{2},j}G_{f,\alpha }^{o}(z_{n})+...  \label{Eq3.6}
\end{eqnarray}
where $z_{n}={(2n+1)i\pi }/{\beta }$ represent the Matsubara
frequencies with  $n$ being any integer.
\begin{equation}
G_{f,\alpha }^{o}(z_{n})=-D_{\alpha }/{(z_{n}-\varepsilon _{f,\alpha})}
\label{bare}
\end{equation}
is the $f$ bare cumulant GF, and in the X-boson the parameter $D_{\alpha }=\langle X_{oo} \rangle +
\langle X_{\alpha\alpha }\rangle$ must be calculated self-consistently.

It is now easy to show that the full calculation of the CHA that includes
the hopping between the $f$-electrons is described by the same diagrams as
the case with $H_{f}^{\prime }=0$, provided that we replace the contribution
$G_{f,0\sigma }^{o}(z_{n})$ of each local vertex with the $\widetilde{G}%
_{j\prime \alpha \prime ;j\alpha }^{ff}(z_{n})$ in Eq. (\ref{Eq3.6}), and
also sum over both internal sites $j$ and $j^{\prime }$, because now they
are not necessarily equal (we do not sum over $j$ or $j^{\prime }$ when they
are external vertices).

When the hopping between localized electrons is included, it is convenient
to first consider the transformation of $\widetilde{G}_{j\prime \alpha
\prime ;j\alpha }^{ff}(z_{n})$ to reciprocal space. Employing Eq. (\ref%
{Eq2.5}) a factor $\exp [\mathbf{k}_{i}.\mathbf{R}_{i}]\ $\ ($\exp [-\mathbf{%
k}_{j}.\mathbf{R}_{j}]$) appears for each internal $X_{i,0\sigma }^{\dagger%
} $($X_{j,0\sigma }$) in $\widetilde{G}_{j\prime \alpha \prime ;j\alpha%
}^{ff}(z_{n})$, while its Fourier transform to reciprocal space provides the
corresponding factors for the two $X$ operators at the end points. Employing
Eq. (\ref{Eq2.5}), one then obtains
\begin{eqnarray}
&\widetilde{G}_{\mathbf{k}\prime \alpha \prime ;\mathbf{k}\alpha
}^{ff}(z_{n})=\delta _{\alpha \prime ,\alpha }\ \delta _{\mathbf{k}\prime,\mathbf{k}}\times \nonumber \\
&\lbrack G_{f,\alpha }^{o}(z_{n})-G_{f,\alpha }^{o}(z_{n})\overline{E}_{\mathbf{k\alpha}}
G_{f,\alpha }^{o}(z_{n})-\nonumber \\
&G_{f,\alpha }^{o}(z_{n})\overline{E}_{\mathbf{k\alpha}}G_{f,\alpha }^{o}(z_{n})
\overline{E}_{\mathbf{k\alpha}}G_{f,\alpha }^{o}(z_{n})+...]\ ,  \label{Eq3.13}
\end{eqnarray}
and then
\begin{equation}
\widetilde{G}_{\mathbf{k}\prime \alpha \prime ;\mathbf{k}\alpha
}^{ff}(z_{n})=\delta _{\alpha \prime ,\alpha }\ \delta _{\mathbf{k}\prime ,%
\mathbf{k}}\ \frac{-{D_{\alpha }}}{z_{n}-\varepsilon _{f,\alpha }-{D_{\alpha
}}\ \overline{E}_{\mathbf{k\alpha}}}.  \label{Eq3.14}
\end{equation}%
The $\delta _{\mathbf{k}\prime ,\mathbf{k}}$ follows from the invariance
against lattice translation of $H_{f}^{\prime }$, (i.e. $t_{i,j}=t_{i-\ell
,j-\ell }$). The $\widetilde{G}_{\mathbf{k}\prime \alpha \prime ;\mathbf{k}%
\alpha }^{ff}(z_{n})$ coincides with the GF of a band of free electrons with
energies $\varepsilon _{f,\alpha }-{D_{\alpha }}\ \overline{E}_{\mathbf{k\alpha}}$%
, except for the ${D_{\alpha }}$ in the numerator. As in the absence of
hopping between $f$-electrons, ${D_{\alpha }}$ describes the effect of the
correlation between these electrons. The GF for the CHA approximation in the
presence of $f$-electron hopping is now easily obtained in reciprocal space
when we notice that the $\exp [\pm \mathbf{k}.\mathbf{R}]$ associated with
the end points of the $\widetilde{G}_{j\prime \alpha \prime ;j\alpha
}^{ff}(z_{n})$, which are provided by the definition of the Fourier transform when
they are external vertices, and by the $V_{j,\mathbf{k},\sigma }=(1/\sqrt{%
N_{s}})V_{\sigma }(\mathbf{k})\exp {(i\mathbf{k}.\mathbf{R}_{j})}$ when they
are connected by hybridization edges to the conduction vertices. In this way,
all the GFs of the CHA in the presence of hopping between $f$-electrons are
then given by:

\begin{equation}
G_{\mathbf{k}{\sigma},\alpha}^{ff}(z_{n})=\frac{-D_{\alpha } \left(
z_{n}-\varepsilon_{\mathbf{k}\sigma }\right) }{\left(z_{n}-E^{f}_{\mathbf{k}%
\alpha}\right)\left(z_{n}-\varepsilon _{\mathbf{k}\sigma}\right) -|V_{\sigma
\alpha}(\mathbf{k})|^{2}D_{\alpha }},  \label{Eq3.15}
\end{equation}
\begin{equation}
G_{\mathbf{k}{\sigma}, \alpha}^{cc}(z_{n})=\frac{-\left(z_{n}-E^{f}_{\mathbf{%
k}\alpha}\right)} {\left(z_{n}-E^{f}_{\mathbf{k}\alpha}\right)\left(z_{n}-%
\varepsilon _{{k}\sigma}\right) -|V_{\sigma \alpha}(\mathbf{k}%
)|^{2}D_{\alpha }},  \label{Eq3.16}
\end{equation}
\begin{equation}
G_{\mathbf{k}{\sigma}, \alpha}^{fc}(z_{n})=\frac{-\ D_{\alpha }V_{\sigma
\alpha}(\mathbf{k})}{\left( z_{n}-E^{f}_{\mathbf{k}\alpha} \right) \left(
z_{n}-\varepsilon _{\mathbf{k}\sigma }\right) -|V_{\sigma \alpha}(\mathbf{k}%
)|^{2}D_{\alpha }}.  \label{Eq3.17}
\end{equation}
\noindent with $E_{\mathbf{k}\alpha}^{f}=\tilde{\varepsilon}_{f}+D_{\alpha}
\overline{E}_{k}$.



\begin{thebibliography}{99}
\bibitem{Doniach1977} {Doniach S 1997 {\it Physica B} {\bf 91} 231}
\bibitem{Coqblin96} {Coqblin B, Arispe J, Iglesias J R, Lacroix C and Le Hur K 1996 
{\it J. Phys. Soc. Jpn.} {\bf 65} 64}
\bibitem{Coqblin97} {Iglesias J R, Lacroix C and Coqblin B 1997 {\it Phys. Rev. B} {\bf 56} 11820}
\bibitem{Coleman84} {Coleman P 1984 {\it Phys. Rev. B} {\bf 29} 3035}
\bibitem{Aeppli92}{Aeppli G and Fisk Z, 1992 {\it Comments Condens. Matter Phys.} {\bf 16} 155 }
\bibitem{Coleman15} Coleman P 2015 Heavy Fermions and the Kondo Lattice: a $21$st Century Perspective arXiv:1509.05769v1 [cond-mat.str-el]
\bibitem{ZFisk12} {Kim D J, Grant T and Fisk Z 2012 {\it Phys. Rev. Lett.} {\bf 109} 096601}
\bibitem{ZFisk13} {Kim D J, Xia J and Fisk Z 2014 {\it Nature Materials} {\bf 13}, 466}
\bibitem{Mignot05} {Mignot J M, Alekseev P A, Nemkovski K S, Regnault L P, Iga F, and Takabatake T 2005{\it Phys. Rev. Lett.} {\bf 94} 247204}
\bibitem{Alekseev09} {Alekseev P A, Lazukov V N, Nemkovskii K S and Sadikov I P 2010 {\it Journal of Experimental and Theoretical Physics} {\bf 111} 285}
\bibitem{Park98} {Park J G, Adroja D T, McEwen K A, Bi Y J and Kulda J 1998 {\it Phys. Rev. B}
\bf{58} 3167}
\bibitem{Sato05} {Sato T J, Kadowaki H, Yoshizawa H, Ekino T, Takabatake T, Fujii H, Regnault L P and Isikawa Y 1995 {\it J. Phys. Condens. Matter} {bf 7} 8009}
\bibitem{Barla05a} {Barla A, Sanchez J P, Derr J, Salce B, Lapertot G, Flouquet J, Doyle B P, Leupold O, Ruffer R, Abd-Elmeguid M M and Lengsdorf R 2005  {\it J. Phys.: Condens. Matter} {\bf 17} S837}
\bibitem{Barla05b} {Barla A, Derr J, Sanchez J P, Salce B, Lapertot G, Doyle B P,  Ruffer R, Lengsdorf R, Abd-Elmeguid M M and Flouquet J 2005 {\it Phys. Rev. Lett.}  {\bf 94} 166401}
\bibitem{Derr08} {Derr J, Knebel G, Braithwaite D, Salce B, Flouquet J, Flachbart K, Gabani S and Shitsevalova N 2008 {\it Phys. Rev. B} \bf{77} 193107}
\bibitem{Paglione16} {Butch N P, Paglione J, Chow P, Xiao Y, Marianetti C A, Booth C H and Jeffries J R 2016 {\it Phys. Rev. Lett.} {\bf 116} 156401}
\bibitem{FengLu13} {Lu F, Zhao J Z, Weng H, Fang Z and Dai X 2013 {\it Phys. Rev. Lett.} {\bf 110} 096401}
\bibitem{Gong15} {Kang C J, Kim J, Kim K, Kang J, Denlinger J D and Min B I 2015 
{\it J. Phys. Soc. Jpn.} {\bf 84} 024722}
\bibitem{Dzero13} {Alexandrov V, Dzero M and Coleman P 2013 {\it Phys. Rev. Lett.} {\bf 111} 226403}
\bibitem{Vojta15} {Baruselli P P and Vojta M 2015 {\it Phys. Rev. Letters} {\bf 115} 156404}
\bibitem{Ikeda96} {Ikeda H and  Miyaki K 1996 {\it J.  Phys. Soc. Jpn.} {bf 65} 1769}
\bibitem{Nakamura96} {Nakamura K I, Kitaoka Y, Asayama K, Takabatake T, Nakamoto G, Tanaka H and Fujii H 1996 {\it Phys. Rev. B} {\bf 53} 6385}
\bibitem{JuanaMoreno2000} {Moreno J and Coleman P 2000 {\it Phys. Rev. Lett.} {\bf 84} 342}
\bibitem{Dzero12} {Dzero M, Sun K, Galitski V  and Coleman P 2010  {\it Phys. Rev. Lett.}  {\bf 104} 106408; Dzero M, Sun K, Coleman P and Galitski V 2012 {\it Phys. Rev. B} {\bf 85} 045130}
\bibitem{Tran12} {Tran M T, Takimoto T  and Kim K S 2012 {\it Phys. Rev. B} {\bf 85} 125128}
\bibitem{Sigrist14} {Legner M, Rueg A and Sigrist M 2014 {\it Phys. Rev. B} {\bf 89} 085110}
\bibitem{Xboson} {Franco R, Figueira M S and Foglio M E 2002 {\it Phys. Rev. B} {bf 66} 045112}
\bibitem {Hubbard5}{Hubbard J 1966 {\it Proc. R. Soc. London Ser A} {\bf296} 82}
\bibitem{FFM}{Figueira M S, Foglio M E and Martinez G G 1994 {\it Phys. Rev. B} {\bf 50} 17933}
\bibitem{Werner13} {Werner J and Assaad F F 2013 {\it Phys. Rev. B} {\bf 88} 035113}
\bibitem{Doniach}  {Doniach S and Sondheimer E H 1974 {Green's Functions for
Solid State Physicists (Benjamin, New York)}
\bibitem{Edwin2014} Ramos E,  Franco R , Silva-Valencia J, Foglio M E and Figueira M S 2014 {\it Journal of Physics: Conference Series} {\bf 568}  052007}
\bibitem{BenHur2000} {Bernhard B H, Lacroix C, Iglesias J R and  Coqblin B 2012 {\it Phys. Rev. B}
{\bf 61}, 441}
\bibitem{Takimoto11}{Takimoto T 2011 {\it J. Phys. Soc. Jpn.} {\bf 80} 123710}
\end{thebibliography}
\end{document}